\documentclass[11pt,letterpaper]{article}
\pdfoutput=1
\usepackage{amscd}
\usepackage{amsmath, bbm} 
\usepackage{gensymb}
\usepackage{slashed}

\usepackage{jheppub} % for details on the use of the package, please

\title{Gauging the accidental symmetries of the Standard Model, and implications for the flavour anomalies}

\author[a]{Wolfgang Altmannshofer,}
\author[b]{Joe Davighi,}
\author[c,d]{and Marco Nardecchia}
\affiliation[a]{Santa Cruz Institute for Particle Physics, University of California, Santa Cruz, CA 95064, USA}
\affiliation[b]{DAMTP, University of Cambridge, Wilberforce Road, Cambridge, 
CB3 0WA, United Kingdom}
\affiliation[c]{INFN - Sezione di Roma, Piazzale Aldo Moro 2, 00185, Roma, Italy}
\affiliation[d]{Sapienza - Universit\'a di Roma, Piazzale Aldo Moro 2, 00185, Roma, Italy}

\emailAdd{waltmann@ucsc.edu}
\emailAdd{jed60@cam.ac.uk}
\emailAdd{marco.nardecchia@cern.ch}

\abstract{We explore the possibility that lepton family numbers and baryon number are such good symmetries of Nature because they are the global remnant of a spontaneously broken gauge symmetry. An almost arbitrary linear combination of these symmetries (together with a component of global hypercharge) can be consistently gauged, if the Standard Model (SM) fermion content is augmented by three chiral SM singlet states.  Within this framework of $U(1)$ extensions of the SM one generically expects flavour non-universality to emerge in the charged leptons, in such a way that naturally prevents lepton flavour violation, by aligning the mass and weak eigenbases. For quarks, all the SM Yukawa couplings responsible for their observed masses and mixings arise at the renormalisable level.
We perform fits to show that models in this class can explain $R_{K^{(\ast)}}$ and the other neutral current $B$ anomaly data if we introduce a heavy vector-like quark to mediate the required quark flavour violation, while simultaneously satisfying other constraints from direct $Z^\prime$ searches at the LHC, $B_s$ meson mixing, a number of electroweak precision observables, and neutrino trident production. Within this symmetry-motivated framework of models, we find interesting implications for the flavour anomalies; notably, any axial couplings of the $Z^\prime$ to electrons and muons must be flavour universal, with the flavour universality violation arising solely from the vector-like couplings. We also comment on the generation of neutrino masses in these models. 
}

\begin{document} 
\maketitle
\flushbottom

\section{Introduction} \label{intro}

The renormalisable Standard Model (SM) lagrangian possesses a number of accidental continuous global symmetries, namely baryon number symmetry, $U(1)_B$, and three individual lepton number symmetries, $U(1)_e$, $U(1)_\mu$, and $U(1)_\tau$. These accidental symmetries are a great success of the SM, since they appear to be extremely good symmetries of Nature that are borne out by (almost) all particle physics experiments to date. The bounds on baryon number-violating processes, most famously from proton decay~\cite{PhysRevD.98.030001}, and on lepton flavour violating (LFV) processes such as $\mu \rightarrow e \gamma$, $\mu \rightarrow eee$, and $\tau\rightarrow \mu\mu\mu$, are very strong~\cite{PhysRevD.98.030001}, and are forecast to strengthen further at future experiments~\cite{Calibbi:2017uvl}. There is only one hint from the world of particle physics to suggest that any of these accidental symmetries are in fact broken, and that is the inference of non-zero neutrino masses from the observation of neutrino oscillations.\footnote{Of course, the violation of baryon number in fundamental interactions is ultimately essential if we are to understand the matter-antimatter asymmetry, and thus the evolution of structure in our Universe.}

This therefore begs the question: {\em why} are these such good symmetries of Nature? As soon as we begin to write down higher-dimension operators in the SM effective field theory (SMEFT), we find that these accidental symmetries of the renormalisable SM lagrangian are broken. The lepton number symmetries are broken already at dimension five, by the inclusion of a Weinberg operator of the form $(LH)^2/\Lambda$, where $\Lambda$ denotes the cut-off scale of the SMEFT. Baryon number is broken at dimension six by four-fermion operators such as $QQQL/\Lambda^2$.

Thus, the first possibility is that $U(1)_B$ and $U(1)_{L_i}$ are such good symmetries simply because the cut-off $\Lambda$ of the SMEFT is very high with respect to energy scales $E$ probed by current particle physics experiments, such as the LHC, so that even dimension five operators are suppressed by a factor $E/\Lambda \ll 1$. While such a scenario certainly remains a viable possibility, it is a rather pessimistic one for particle physicists to swallow, for it implies, by and large, that {\em any} new physics beyond the Standard Model (BSM) is likely lying out of reach of current colliders.

The second and more intriguing possibility is that, in some extended BSM theory that arises due to new physics at a scale $\Lambda_{\text{LFUV}}$, these global symmetries remain accidental at the level of the renormalisable lagrangian. If this is the case, one would expect a natural separation between the scale $\Lambda_{\text{LFUV}}$ of those higher-dimension operators which respect these global symmetries, and the scale  $\Lambda_{\text{LFV}}\gg \Lambda_{\text{LFUV}}$ of those operators which violate them (where the reason for these names shall soon be apparent).\footnote{
One might try to pursue a middle way between these two options, and suggest that there 
exists such a hierarchy of scales in the SMEFT, with $U(1)_B$- or $U(1)_{L_i}$-violating operators being suppressed by some cut-off scale $\Lambda_{\text{LFV}} \gg \Lambda_{\text{LFUV}}$, not because of any underlying gauge symmetry, but simply because of small couplings in the EFT (in other words, one chooses to forgo the expectation from na\"ive dimensional analysis (NDA), and the `second scale' really signifies the existence of very small couplings $c$ such that the ratio $\Lambda_{\text{LFUV}}/c\sim \Lambda_{\text{LFV}}  \gg \Lambda_{\text{LFUV}}$).
This approach is at least consistent, in the sense that the smallness of the couplings $c$ can be radiatively stable in the low-energy EFT.}
In such a scenario, the true cut-off $\Lambda_{\text{LFUV}}$ of the SMEFT can in fact be brought much lower in energy than the scale at which $U(1)_B$ and $U(1)_{L_i}$ are violated. While this second option is certainly favoured by the biases of optimistic model builders, until recently there has been no physics case for favouring either one of these two hypotheses over the other.

Fortunately, this {\em impasse} is now being challenged by the recent emergence of a set of intriguing discrepancies between SM predictions and experimental measurements, in the neutral current decays of $B$ mesons, which would seem to favour the second of the two possibilities just discussed. For example, the ratio of branching ratios $R_{K^{(\ast)}} \equiv BR(B \rightarrow K^{(\ast)} \mu^+ \mu^-) / BR(B\rightarrow K^{(\ast)} e^+ e^-)$ is equal to unity in the SM to the percent level, for di-lepton invariant mass squared bin $q^2 \in [1.1,6]$ GeV$^2$, but LHCb has measured~\cite{Aaij:2017vbb,Aaij:2019wad} $R_K=0.846^{+0.060}_{-0.054}{}^{+0.016}_{-0.014}$ and $R_{K^{\ast}}=0.69^{+0.11}_{-0.07}\pm0.05$ in this $q^2$ bin, corresponding to deviations from the SM by approximately $2.5\sigma$ each. LHCb has also measured $R_ {K^\ast}=0.66^{+0.11}_{-0.07}\pm 0.03$ for the low momentum bin $q^2\in \left[0.045,1.1\right]\mathrm{GeV}^2$, which is again about $2.5\sigma$ under the SM prediction~\cite{Aaij:2017vbb}.
There are further notable discrepancies with the SM predictions in measurements of $BR(B_s\rightarrow \mu\mu)$ \cite{Aaboud:2018mst,Chatrchyan:2013bka,CMS:2014xfa,Aaij:2017vad}, and in $B\rightarrow K^\ast \mu^+\mu^-$ angular observables such as $P_5^\prime$~\cite{Aaij:2013qta,Aaij:2015oid,CMS:2017ivg,Aaboud:2018krd}.
For a comprehensive survey of these anomalies in the decays of neutral $B$ mesons, which we henceforth refer to collectively as the `neutral current $B$ anomalies' (NCBAs), see {\em e.g.} \cite{Aebischer:2019mlg}. 

Taken together, these measurements point towards lepton flavour universality violation (LFUV) between $e$ and $\mu$.\footnote{There are also experimental hints of LFUV between $\tau$ and light leptons ($e/\mu$) in charged current $B$ meson decays to $D^{(\ast)}\tau\nu$ ~\cite{Lees:2012xj,Lees2013,Aaij:2015yra,Huschle:2015rga,Sato:2016svk,Hirose2017,Hirose:2017dxl}. However, we shall only consider the neutral current anomalies in this paper.} But crucially, there is no evidence for LFV. Thus, these tantalising first hints of BSM physics at the LHC seem to respect the accidental global symmetries of the SM, despite violating lepton flavour universality. The magnitude of the deviations from the SM in the various NCBAs suggests there is new physics at a cut-off scale of order $\Lambda_{\text{LFUV}} \sim 30$ TeV ~\cite{Alguero:2019ptt,Alok:2019ufo,Ciuchini:2019usw,Aebischer:2019mlg,Kowalska:2019ley,Arbey:2019duh} or thereabouts (most pessimistically, constraints from perturbative unitarity imply the new physics scale cannot be larger than about $\Lambda_{\text{LFUV}} \sim 80$ TeV ~\cite{DiLuzio:2017chi}).
This energy scale is probed at the LHC by measurements of many rare flavour observables, not just those mentioned above, and there are no other signs of new physics at this scale, and certainly no evidence for the violation of $U(1)_B$ or $U(1)_{L_i}$. In this way, if the NCBAs persist in the wake of future measurements (from LHCb, Belle II, and others~\cite{Albrecht:2017odf}), and if these or any other new anomalies continue to respect $U(1)_B$ and $U(1)_{L_i}$, it would seem to suggest that the SM's global symmetries $U(1)_B$ and $U(1)_{L_i}$ are likely a consequence of some underlying dynamics, such as a gauge symmetry.

If we adopt such an optimistic interpretation of the NCBAs, then which symmetry might one gauge, which can both explain the NCBAs and would also underwrite these global symmetries? At the level of the SMEFT, the NCBA data can be explained by BSM contributions to the following four-fermion operators
\begin{eqnarray}
  {\mathcal L}_{bs\ell\ell} =
  C_L^e(\overline{s_L} \gamma_\rho b_L) 
  (\overline{e_L} \gamma^\rho e_L) +
  C_R^e(\overline{s_L} \gamma_\rho b_L) 
  (\overline{e_R} \gamma^\rho e_R) +\nonumber \\
  C_L^\mu (\overline{s_L} \gamma_\rho b_L) 
  (\overline{\mu_L} \gamma^\rho \mu_L) +
  C_R^\mu (\overline{s_L} \gamma_\rho b_L) 
  (\overline{\mu_R} \gamma^\rho \mu_R), \label{bsll}
\end{eqnarray}
where the SM contribution is $C_L^{e,SM}=C_L^{\mu,SM} \simeq 8.64/(36 \text{~TeV})^2$ and $C_R^{e,SM} \simeq C_R^{\mu,SM}=-0.18/(36 \text{~TeV})^2$ (borrowing the numerics from Ref.~\cite{DAmico:2017mtc}), which is due to one-loop $W$ exchange. Global fits to the data favour a significant BSM contribution to $C_L^\mu$, and are consistent with non-zero BSM contributions to the other three operators~\cite{Alguero:2019ptt,Alok:2019ufo,Ciuchini:2019usw,Aebischer:2019mlg,Kowalska:2019ley,Arbey:2019duh}. One simple possibility is that all these operators receive BSM contributions due to the tree-level exchange of a heavy $Z^\prime$ vector boson, which couples to muons and (possibly) electrons, in addition to possessing a flavour violating coupling to $b\bar{s}$. While the NCBAs can also be explained by tree-level leptoquark exchange or by various loop-induced processes, in this paper we shall restrict our attention to $Z^\prime$ models, in which the $Z^\prime$ arises from a spontaneously broken, flavour-dependent $U(1)$ gauge symmetry, which we shall denote henceforth by $U(1)_X$.

There are many possible $U(1)_X$ symmetry groups that can be gauged in order to explain the NCBAs, and the literature on such models has grown vast  (see {\em e.g.} Refs.~\cite{Gauld:2013qba,Buras:2013qja,Altmannshofer:2014cfa,Crivellin:2015mga,Crivellin:2015lwa,Sierra:2015fma,Crivellin:2015era,Celis:2015ara,Altmannshofer:2015mqa,Allanach:2015gkd,Falkowski:2015zwa,Becirevic:2016zri,Boucenna:2016wpr,Ko:2017lzd,Alonso:2017bff,Kamenik:2017tnu,Alonso:2017uky,Faisel:2017glo,PhysRevD.97.115003,Bian:2017xzg,PhysRevD.97.075035,Bhatia:2017tgo,Duan:2018akc,Allanach:2018odd,Allanach:2018lvl} for an incomplete list). In this work, we shall make a number of simplifying assumptions before arriving at a new framework of SM$\times U(1)_X$ theories in which the protection of the SM's accidental symmetries is built-in. Firstly, we explore only $U(1)_X$ charge assignments which {\em allow a SM-like Yukawa sector for the quarks}, at the renormalisable level. Within such an approach we make no attempt to shed light upon the peculiar flavour structure of the SM, which features large hierarchies in fermion mass parameters and the quark mixing angles. On the other hand, we avoid having to delve more deeply into the model building to explain the origin of Yukawa couplings if they are to be banned at the renormalisable level, as they are in {\em e.g.} Refs.~\cite{Allanach:2018lvl,Davighi:2019jwf,Allanach:2019iiy}. 

In the lepton sector, we shall require that {\em only the diagonal charged lepton Yukawa couplings are $U(1)_X$-invariant}, and thus present in the renormalisable lagrangian. This is achieved (given only a single Higgs, possibly charged under $U(1)_X$) by requiring the different family leptons have different $U(1)_X$ charges, but that the differences between the left-handed and right-handed lepton charges are family universal. Such a charge assignment automatically implies LFUV, since the $Z^\prime$ couples differently to each lepton family. Furthermore, in banning all the off-diagonal charged lepton Yukawa couplings (at the renormalisable level), the charged lepton mass eigenbasis is aligned with the weak eigenbasis,\footnote{The PMNS mixing must come entirely from the neutrino sector of the model.} and hence no lepton flavour violating neutral currents will be induced in the physical mass basis. Thus, the very same $U(1)_X$ gauge symmetry which introduces LFUV also prevents LFV, in opposition to the claims made in Ref.~\cite{Glashow:2014iga}.\footnote{A salient point here is that even when a gauge symmetry is spontaneously broken, this does not imply there is not a global part of that gauge symmetry that remains unbroken (when considering only its action on fermions). This is the case, for example, with the global part of the hypercharge symmetry of the SM after electroweak symmetry breaking.} We must also require our $U(1)_X$ symmetry be {\em free of all gauge anomalies}, including mixed and gravitational anomalies.\footnote{We note in passing that in many $Z^\prime$ models which are necessarily only low-energy EFTs (for example, because they do not permit a gauge-invariant Yukawa sector), it is not essential to cancel gauge anomalies at low energies; it might be possible in such models for anomaly cancellation to be restored in the high-energy theory by {\em e.g.} `integrating in' a set of heavy chiral fermions, or by the Green Schwarz mechanism~\cite{Coriano:2007fw,Coriano:2007xg}. In our setup, we seek to embed our $Z^\prime$ in a fully renormalisable extension of the SM, and thus anomaly cancellation is for us an essential requirement.} In these senses, the SM$\times U(1)_X$ theories we shall define may be regarded as technically `complete', minimal extensions of the SM, albeit with an unexplained flavour structure as in the SM itself.

What is the most general SM$\times U(1)_X$ theory which satisfies these criteria? In the spirit of minimality, we want to add as few chiral states as possible beyond those of the SM. The most minimal option, of course, is to add no BSM states at all (beyond the $Z^\prime$, and a heavy scalar field $\Phi$ introduced to spontaneously break $U(1)_X$). In this case, it has been known for a long time that the only anomaly-free $U(1)_X$ charge assignments consistent with our criteria correspond to gauging $L_i-L_j$, the difference of any pair of lepton numbers~\cite{PhysRevD.44.2118}.\footnote{In fact, while this statement is `common lore' amongst physicsts, we shall discover in \S \ref{framework} that, under these conditions, one may gauge a slightly larger symmetry, of the form $(L_i-L_j)+aY$, where $Y$ denotes hypercharge and $a$ is any rational number.} The particular choice of gauging $L_\mu - L_\tau$ offers a compelling $Z^\prime$ explanation of the NCBAs (after flavour violating couplings to quarks are introduced, for example through effective non-renormalisable interactions), which moreover underwrites the SM's global lepton number symmetries with a gauge symmetry in the manner described above~\cite{Altmannshofer:2014cfa,Altmannshofer:2015mqa}. The $Z^\prime$ boson of such a model can even mediate interactions with a dark sector; connecting the NCBAs with the dark matter problem through such a gauged $L_\mu-L_\tau$ leads to a highly-testable model with rich phenomenology~\cite{Altmannshofer:2016jzy}.

The next minimal option is to introduce a `dark sector' of SM singlet chiral fermion states, which are charged only under $U(1)_X$. We shall here consider augmenting the SM matter content with up to three such states. Since the extra states are chiral only with respect to the $U(1)_X$ symmetry, they do not spoil the cancellation of anomalies in the SM gauge sector. Nonetheless, these states contribute to two out of the six anomaly coefficients involving $U(1)_X$, and their inclusion can thus be used to cancel two anomaly coefficients that would otherwise be non-zero, thereby opening up a wider space of anomaly-free charge assignments beyond just $L_i-L_j$. Within this setup, there is a four-parameter family of such anomaly-free $U(1)_X$ symmetries that can be gauged, generated by
\begin{equation}\label{U1X}
T_X = a_Y \, T_{Y} - a_e \, T_{B/3-L_e} - a_{\mu} \, T_{B/3-L_{\mu}} - a_{\tau} \, T_{B/3-L_{\tau}}, 
\end{equation}
where $a_e, a_{\mu}, a_{\tau}$ and $a_y$ are rational coefficients, and $T_Y$ denotes the generator of (global) hypercharge. We may rewrite this generator in terms of the generators of the accidental global symmetries of the SM, as
\begin{equation}\label{U1X b}
T_X = \sum_i a_i T_{L_i} - \left( \frac{a_e + a_{\mu}+a_{\tau}}{3} \right) \, T_B + a_Y \, T_{Y},
\end{equation}
where here $a_i\in \{a_e,a_\mu,a_\tau\}$. This intriguing result, which we shall review in \S \ref{framework}, was originally derived in Ref.~\cite{Salvioni:2009jp}, in which the three chiral dark states were interpreted as right-handed neutrinos.\footnote{A three-parameter subset of these gauge symmetries, in which it was assumed that $a_Y=0$, was also considered in Ref.~\cite{Araki:2012ip} (and subsequently in Refs.~\cite{heeck2014neutrinos,Asai:2019ciz}, for example), with the goal of explaining the neutrino masses and mixing parameters. In our setup, we shall {\em not} interpret the three chiral dark states as right-handed neutrinos; rather, we shall suggest that neutrino masses may arise from a rather different mechanism - see Appendix \ref{neutrinos}. Various other models in which more than one of the global SM symmetries is gauged have also been considered in the literature, {\em e.g.} in Refs.~\cite{FileviezPerez:2010gw,Perez:2014qfa}.} 
To summarise, if one allows the addition of three SM singlet states to `soak up' anomalies, then the {\em most general anomaly-free $U(1)_X$ charge assignment for the SM fermions}, which allows a fully generic quark Yukawa sector and a strictly diagonal charged lepton Yukawa matrix, corresponds to {\em gauging an almost arbitrary linear combination of the (otherwise accidental) global symmetries of the SM} (including the `global part' of hypercharge symmetry).  The word `almost' indicates an important caveat, that the linear combination in (\ref{U1X b}) is not entirely arbitrary; in particular, the component of baryon number is fixed by the components of each of the lepton numbers. This means, for example, that one cannot gauge $B+L$, which is  of course anomaly-full. As long as the coefficients $\{a_e,a_\mu,a_\tau\}$ are all different, gauging such a symmetry implies LFUV, while preventing LFV.

In the rest of this paper, we develop $Z^\prime$ models based on gauging $U(1)_X$ symmetries in the family defined by (\ref{U1X}), and explore their phenomenology. In some sense, these models provide a wide generalisation of the $L_\mu-L_\tau$ model, and so several aspects of our setup shall be borrowed from Ref.~\cite{Altmannshofer:2014cfa} (such as the mechanism for generating quark flavour violating couplings of the $Z^\prime$ to $b\bar{s}$). We show in \S \ref{anomalies} that these models make a distinctive prediction for the structure of the NCBAs, which may be decomposed into a flavour non-universal vector coupling to leptons, plus a flavour-universal axial component (if $a_Y\neq 0$). Furthermore, if this axial component of the anomaly is non-vanishing, the $Z^\prime$ inevitably must acquire tree-level couplings to valence quarks. We extract new global fits to the NCBA data using \texttt{flavio}~\cite{Straub:2018kue}, in terms of the parameters of our model. We identify a physically motivated benchmark scenario in order to interpret these multi-dimensional fits, by fixing $a_\mu = 1$ and $a_\tau=0$, and compute other important phenomenological bounds in \S \ref{pheno} at this benchmark point in our parameter space, before concluding.

Our goal in this paper is not to explore the phenomenology of these theories completely; rather, we are content to demonstrate that our framework leads to rich phenomenology, as hinted at above, and that there is interesting allowed parameter space, thereby laying the groundwork for possible future studies.

In Appendix \ref{neutrinos} we discuss how light neutrino masses might naturally arise within our framework, which induce lepton flavour violation at the higher scale $\Lambda_{\text{LFV}}$. This involves a more in-depth examination of the dark sector of the theory, for the more enthusiastic of our readers.

\section{A new framework for LFUV without LFV} \label{framework}

We consider an extension of the SM by a flavour-dependent $U(1)_X$ gauge symmetry, under which all SM fields plus three SM singlet chiral fermions (which we will denote by $\nu_R^i$) may be charged {\em a priori}. The $U(1)_X$ symmetry will be spontaneously broken by the vacuum expectation value (vev) $v_\Phi$ of a SM singlet scalar field $\Phi$, at around the TeV scale, leading to a heavy $Z^\prime$ gauge boson. We require:
\begin{enumerate}
\item That the assignment of $U(1)_X$ charges is anomaly-free.
\item That all Yukawa couplings in the quark sector are permitted at the renormalisable level.
\item That the flavour-dependent $U(1)_X$ gauge symmetry protects each individual lepton number symmetry, and thus forbids lepton flavour violating (LFV) processes. This is achieved by aligning the mass eigenbasis of charged leptons with the weak eigenbasis. Thus, we require that only the diagonal elements are present in the charged lepton Yukawa matrix (at the renormalisable level), which in turn requires lepton flavour universality violation (LFUV) between all three families.
\end{enumerate}
We write the fermion fields (including the three SM singlets) as the following representations of the gauge group $SU(3) \times
SU(2)_L \times U(1)_Y \times U(1)_X$: 
$$
q^i_L \sim (\mathbf{3}, \mathbf{2}, 1/6, \hat{Q}_q), \quad
u^i_R \sim (\mathbf{3}, \mathbf{1}, 2/3, \hat{Q}_u), \quad
d^i_R \sim (\mathbf{3}, \mathbf{1}, -1/3, \hat{Q}_d),
$$
$$
\ell^i_L \sim (\mathbf{1}, \mathbf{2}, -1/2, \hat{Q}_{\ell^i}), \quad
e^i_R \sim (\mathbf{1}, \mathbf{1}, -1, \hat{Q}_{e^i}), \quad
\nu^i_R \sim (\mathbf{1}, \mathbf{1}, 0, \hat{Q}_{\nu^i}),
$$
where the index $i \in \{1,2,3\}$ denotes the family. Our assumptions regarding fully generic quark Yukawa couplings imply the quark charges under $U(1)_X$ are flavour universal, whereas our mechanism for preventing LFV requires the lepton charges are flavour {\em non}-universal. The scalar sector of the theory contains the Higgs and the scalar $\Phi$, which carry the representations 
$$
H \sim (\mathbf{1}, \mathbf{2}, -1/2, \hat{Q}_H), \quad \Phi \sim (\mathbf{1}, \mathbf{1}, 0, \hat{Q}_\Phi).
$$
The charges under $U(1)_X$, denoted by the set of $\hat{Q}$s, are all assumed to be rational numbers.\footnote{If the charges under $U(1)_X$ are not rational numbers, then such matter content cannot arise from any unified gauge theory with semi-simple gauge group $G$. There are good reasons to assume that such a unified theory ultimately describes the interactions of elementary particles in the ultraviolet.}

In such an extension of the SM by a (flavour-dependent) $U(1)_X$ gauge symmetry, there are six independent anomaly coefficients which must vanish, corresponding to the six (potentially non-vanishing) triangle diagrams involving at least one $U(1)_X$ gauge boson: $SU(3)^2\times U(1)_X$, $SU(2)_L^2\times U(1)_X$, $U(1)_Y^2\times U(1)_X$, $U(1)_Y\times U(1)_X^2$, $U(1)_X^3$, and the mixed gauge-gravity anomaly involving $U(1)_X$. Thus, anomaly cancellation implies a system of six non-linear equations over the eighteen chiral fermion charges listed above.\footnote{One may of course add arbitrary scalars or vector-like fermions without affecting anomaly cancellation.} By rescaling the gauge coupling, we may take these rational charges to be integers, and then apply arithmetic methods (such as those introduced in Ref.~\cite{Allanach:2018vjg}) to solve the resulting system of non-linear Diophantine equations.

In fact, it shall turn out that because of the heavy restrictions we are enforcing from the Yukawa sector, the subspace of anomaly-free solutions that we are interested in is essentially picked out by a strictly {\em linear} system of equations, and so may be extracted using basic linear algebra; thus, employing the Diophantine methods outlined in Ref.~\cite{Allanach:2018vjg} in this case would be something like overkill.

We begin by enforcing the constraints on the charges coming from the Yukawa sector. As already noted, requiring a renormalisable quark Yukawa sector implies that $q_L$, $u_R$, and $d_R$ have flavour-universal $U(1)_X$ charges, which we denote by $\hat{Q}_q$, $\hat{Q}_u$, and $\hat{Q}_d$. 
Then, requiring renormalisable Yukawas for up and down quarks, and diagonal charged lepton Yukawas in each generation, implies the following five linear constraints (which are all linearly independent) on the remaining ten charges:
\begin{eqnarray}\label{yukawa constraints}
\hat{Q}_q-\hat{Q}_u &=& \hat{Q}_H,\nonumber \\
\hat{Q}_q-\hat{Q}_d &=& -\hat{Q}_H,\nonumber \\
\hat{Q}_{\ell^i}-\hat{Q}_{e^i} &=& -\hat{Q}_H, \qquad i\in\{1,2,3\}.
\end{eqnarray}
We are thus reduced, by these linear constraints, to a five-parameter family of charges. As might be expected, this family of charge assignments corresponds to gauging an arbitrary linear combination of the five accidental global symmetries of the SM Yukawa sector: $U(1)_B$, $U(1)_e$, $U(1)_\mu$, and $U(1)_\tau$, and (the global part of) hypercharge, $U(1)_Y$. But such a charge assignment is not yet anomaly-free, an issue that we shall now remedy.

Firstly, it is helpful to notice that the three SM singlet dark states appear in just two of the six anomaly cancellation equations. Specifically, these are the (linear) gauge-gravity anomaly, whose coefficient is proportional to
\begin{equation}
  A_\mathrm{grav} = 18 \hat{Q}_{q}  - 9 \hat{Q}_{u} - 9 \hat{Q}_{d} + \sum_{i=1}^3 (2 \hat{Q}_{\ell^i}
  -\hat{Q}_{e^i} - \hat{Q}_{\nu^i}), \label{grav} 
\end{equation}
and the (cubic) $U(1)_X^3$ anomaly, with coefficient proportional to
\begin{equation}
  A_\mathrm{cubic} = 18 \hat{Q}^3_{q}- 9 \hat{Q}^3_{u} - 9 \hat{Q}^3_{d} + \sum_{i=1}^3( 2 \hat{Q}^3_{\ell^i} 
  -\hat{Q}^3_{e^i} - \hat{Q}^3_{\nu^i}). \label{cubic} 
\end{equation}
Thus, the charges of the SM fermions {\em on their own} must satisfy the {\em other} four anomaly equations independently. 

Of these, three are linear, and moreover out of these three linear constraints, only one turns out to be linearly-independent from (\ref{yukawa constraints}).\footnote{This additional constraint can be taken to be either the $U(1)_Y^2\times U(1)_X$ anomaly equation or the $SU(2)^2_L\times U(1)_X$ anomaly equation or indeed a linear combination of these two, but {\em not} the $SU(3)^2\times U(1)_X$ anomaly equation.} This additional linear constraint has the effect of fixing the component of baryon number in terms of the other symmetries (in such a way that excludes the gauging of $B+L$, which is well known to be anomaly-full). We are thus reduced by this subset of the anomaly cancellation equations to a four-parameter family of solutions. A particularly suggestive parametrisation for the SM fermion charges, in terms of four rational numbers $a_e, a_{\mu}, a_{\tau}$, and $a_Y$, is recorded in the first ten rows of the following:

\begin{equation} \label{charges}
\begin{array}{c|c|l}
\textrm{Field~} & G_{\text{SM}} & U(1)_X\ \\
\hline
q_L^i & (\mathbf{3},\mathbf{2},1/6) & \hat{Q}_q =  (-a_e-a_\mu-a_\tau)/9+a_Y/6\\
u_R^i & (\mathbf{3},\mathbf{1},2/3) & \hat{Q}_u =  (-a_e-a_\mu-a_\tau)/9+2a_Y/3 \\
d_R^i & (\mathbf{3},\mathbf{1},-1/3) & \hat{Q}_d =  (-a_e-a_\mu-a_\tau)/9-a_Y/3 \\
\hline
\ell_L^1 & (\mathbf{1},\mathbf{2},-1/2) & \hat{Q}_{\ell^1} = a_e - a_Y/2 \\ 
\ell_L^2 & (\mathbf{1},\mathbf{2},-1/2) & \hat{Q}_{\ell^2}  = a_{\mu} - a_Y/2 \\
\ell_L^3 & (\mathbf{1},\mathbf{2},-1/2) & \hat{Q}_{\ell^3}  = a_{\tau} - a_Y/2 \\
e_R^1 & (\mathbf{1},\mathbf{1},-1) & \hat{Q}_{e^1}  = a_e - a_Y \\
e_R^2 & (\mathbf{1},\mathbf{1},-1) & \hat{Q}_{e^2} = a_{\mu} - a_Y \\ 
e_R^3 & (\mathbf{1},\mathbf{1},-1) & \hat{Q}_{e^3} = a_{\tau} - a_Y \\
\hline
H & (\mathbf{1},\mathbf{2},-1/2) & \hat{Q}_H=-a_Y/2 \\
\hline
\nu^1_R & (\mathbf{1},\mathbf{1},0) & \hat{Q}_{\nu^1}=a_e \\
\nu^2_R & (\mathbf{1},\mathbf{1},0) & \hat{Q}_{\nu^2}=a_\mu \\
\nu^3_R & (\mathbf{1},\mathbf{1},0) & \hat{Q}_{\nu^3}=a_\tau \\
\Phi & (\mathbf{1},\mathbf{1},0) & \hat{Q}_{\Phi}=1 \\
Q_{L/R} & (\mathbf{3},\mathbf{2},1/6) & \hat{Q}_Q  = -\hat{Q}_{\Phi} + \hat{Q}_q,
\end{array}
\\[15pt]
\end{equation}
where in the bottom five rows we have also included the states in our `dark sector' here for completeness; there are the three dark chiral fermion states $\nu^i$, which are SM singlets, the scalar $\Phi$ whose role it is to spontaneously break $U(1)_X$ by acquiring a vev, and a vector-like heavy fermion denoted $Q$, which plays an important role in introducing quark flavour violation into the interactions of the $Z^\prime$, which we shall discuss in \S \ref{subsec:QFV}. 

The coefficient of the $U(1)_X^2\times U(1)_Y$ anomaly, which we have not yet considered, is proportional to the quadratic expression
\begin{equation}
A_{\mathrm{quad}}= 3\hat{Q}^2_{q} - 6 \hat{Q}^2_{u} + 3\hat{Q}^2_{d} +   \sum_{i=1}^3( \hat{Q}^2_{e^i}- \hat{Q}^2_{\ell^i} ), 
\label{quad} 
\end{equation}
Somewhat surprisingly (or perhaps not), when we substitute into (\ref{quad}) the general solution parametrised in (\ref{charges}), we find that $A_{\mathrm{quad}}=0$ for any values of $(a_e, a_\mu, a_\tau, a_Y)\in \mathbb{Q}^4$.

At this stage, we turn to the two anomalies which are sensitive to the dark sector chiral fermions $\nu_R^i$, via the anomaly coefficients in (\ref{grav}, \ref{cubic}). If we substitute in the charge assignment in (\ref{charges}), which we have derived as the most general charge assignment for the SM sector fermions (that is consistent with the other constraints from anomaly cancellation and the Yukawa sector), we obtain
\begin{equation}
  A_\mathrm{grav} = a_e+a_\mu+a_\tau - \sum_{i=1}^3 \hat{Q}_{\nu^i}, \qquad \text{and} \qquad A_\mathrm{cubic} = a_e^3+a_\mu^3+a_\tau^3- \sum_{i=1}^3 \hat{Q}^3_{\nu^i}. \label{dark anomalies}
\end{equation}
In the absence of the dark sector fermions ($\hat{Q}_{\nu^i}$=0), the only possible solution\footnote{It is a rather famous theorem in number theory that $a_e^3+a_\mu^3+a_\tau^3=0$, where $a_e$, $a_\mu$, and $a_\tau$ are three rational numbers, implies $a_e a_\mu a_\tau=0$.} to $A_\mathrm{cubic}=0$ is to choose one of $a_e$, $a_\mu$, and $a_\tau$ to be zero, and the other two to sum to zero; $a_Y$ remains unconstrained. This also satisfies $A_\mathrm{grav}=0$, and thus returns anomaly-free solutions of the form $a(L_i-L_j)+bY$. But by introducing the three dark sector fermions, we can absorb {\em any} remaining anomalies in $A_{\mathrm{grav}}$ or $A_{\mathrm{cubic}}$ by ascribing them charges
\begin{equation}
\hat{Q}_{\nu^1}=a_e, \quad \hat{Q}_{\nu^2}=a_\mu, \quad \hat{Q}_{\nu^3}=a_\tau, \label{dark charges}
\end{equation}
or any permutation thereof, which would correspond to the lepton numbers of right-handed neutrinos, with one assigned to each generation.\footnote{We are not claiming that the assignment of the dark sector fermion charges in (\ref{dark charges}) is the {\em only} solution to the pair of equations (\ref{dark anomalies}); rather, we are content that there is always {\em guaranteed} to be a solution to all the anomaly equations with dark sector charges of this form. Indeed, for certain rational values of the triple $(a_e,a_\mu,a_\tau)$ it is known that other non-trivial solutions exist~\cite{Costa:2019zzy}. In particular, if we rescale the gauge coupling (without loss of generality) such that $(a_e,a_\mu,a_\tau)$ and $\{\hat{Q}_{\nu^i}\}$ are all integers, then an algorithmic method has been developed in Ref.~\cite{Costa:2019zzy} for finding all solutions to (\ref{dark anomalies}) for the six numbers $(a_e,a_\mu,a_\tau,\hat{Q}_{\nu^1},\hat{Q}_{\nu^2},\hat{Q}_{\nu^3})$. Examples include $(5,-4,-4,-1,-1,-1)$, $(6,-5,-5,-3,-2,1)$, and $(11,-9,-9,-4,-4,1)$~\cite{Costa:2019zzy}. In the present paper we wish to be able to vary $(a_e,a_\mu,a_\tau)$ freely within the rationals, in order to carry out fits to the NCBA data; in this situation, there do not necessarily exist {\em any} `non-trivial' solutions to (\ref{dark anomalies}) beyond the charge assignment (\ref{dark charges}) that we choose. Nonetheless, we shall reconsider such `non-trivial solutions' when we come to discuss neutrino masses in Appendix~\ref{neutrinos}.} The anomaly-freedom of this charge assignment was originally shown in Ref.~\cite{Salvioni:2009jp}.

The charge assignment in (\ref{charges}) corresponds to gauging a $U(1)_X$ symmetry generated by
\begin{equation}
T_X = a_e T_{L_e}+a_\mu T_{L_\mu}+a_\tau T_{L_\tau} - \left( \frac{a_e + a_{\mu}+a_{\tau}}{3} \right) \, T_B + a_Y \, T_{Y},
\end{equation}
{\em i.e.} an almost arbitrary linear combination of the accidental symmetries of the SM, namely baryon number, the three individual lepton numbers, and the global part of hypercharge (with that linear combination being `orthogonal' to $B+L$). Of course, this is by no means a miracle; the very fact that these quantities are accidental symmetries of the SM lagrangian implies that the Yukawa sector is invariant under precisely these symmetries. What is surprising, at least to us, is that an (almost) arbitrary linear combination of these global symmetries can be made anomaly-free, and thus can be gauged, with only a minimal extension of the SM field content by three chiral SM singlets. 

Provided that $a_e$, $a_\mu$, and $a_\tau$ are all different, in other words that the charge assignment violates lepton flavour universality in all three families, the charged lepton Yukawa matrix will be strictly diagonal; thus, the same symmetry which introduces LFUV simultaneously prevents LFV.

\section{Towards a model for the \texorpdfstring{$B$}{B} anomalies}

In this Section, we develop SM$\times U(1)_X$ gauge theories in this family into phenomenological models capable of explaining the NCBAs. For ease of reference, we summarise the full lagrangian of the model in Eq. (\ref{full lagrangian}).

In order to mediate the flavour-changing neutral current interactions in (\ref{bsll}), we must introduce flavour-changing quark couplings into our framework. Since the $U(1)_X$ gauge symmetry couples universally to the three quark generations, this cannot be achieved simply by the CKM mixing between the weak and mass eigenbases. The same issue afflicts (for example) models based on gauging $L_\mu-L_\tau$ (in which the $Z^\prime$ doesn't couple directly to quarks at all). We shall thus generate quark flavour violation in our model by following a similar procedure to that found in Ref.~\cite{Altmannshofer:2014cfa}, which in turn is based on the `effective operator' approach first suggested in Ref.~\cite{Fox:2011qd}. We will then go on to describe the mass mixing between the neutral gauge bosons, specifically the $Z$ and the $Z^\prime$, that occurs in these theories when $a_Y\neq 0$, for which we largely follow Refs.~\cite{Allanach:2018lvl,Allanach:2019iiy}.

\subsection{Quark flavour violation} \label{subsec:QFV}

We introduce a heavy\footnote{We shall clarify exactly what we mean by `heavy' shortly.} vector-like quark field, denoted $Q$, whose left- and right-handed components both transform in the representation
$$Q_{L/R}\sim (\mathbf{3},\mathbf{2},1/6,\hat{Q}_Q)$$
of the $SU(3)\times SU(2)_L\times U(1)_Y \times U(1)_X$ gauge symmetry. In other words, other than the $U(1)_X$ charge and its vector-like nature, the field $Q$ is a `heavy copy' of the quark doublet field $q_L$ in the SM. Indeed, it shall be convenient to decompose $Q$ into its $SU(2)_L$ components, {\em viz.}
\begin{equation}
Q_L = (u_L^4,d_L^4)^T, \qquad Q_R = (u_R^4,d_R^4)^T,
\end{equation}
where the index anticipates that we shall soon view $u_L^4$ and $d_L^4$ as fourth family quark fields. Because $Q$ is a vector-like fermion, it does not spoil the anomaly cancellation in our setup.
Together with the fields already described above, this completes the field content of our framework of models, as recorded in (\ref{charges}).

Given the quantum numbers of $Q$ and of the SM quark fields, we can write down the following terms in the lagrangian, which result in effective mass terms after both $\Phi$ and the Higgs acquire their vevs:
\begin{equation} \label{mass terms}
\mathcal{L} \supset - \left[ (Y_D)_{ij} \, \overline{q}_L^i d_R^j H^c + (Y_U)_{ij} \, \overline{q}_L^i u_R^j H  \right] + \mathcal{L}_{\text{mix}}, \qquad \mathcal{L}_{\text{mix}} = -m_Q \overline{Q} Q + (Y_{Qi} \, \overline{q}_L^i Q_R \Phi + \textrm{h.c}),
\end{equation} 
where $H^c=(H^+, -H^{0\ast})^T$. The term $-m_Q \overline{Q} Q$ is simply a vector-like mass term for $Q$. The first two terms within the square brackets are the usual Yukawa couplings for the down and up quarks, while the second term in $\mathcal{L}_{\text{mix}}$ leads to mass mixing between the vector-like quark and the SM quarks. Just like the SM Yukawa couplings $(Y_D)_{ij}$, the $Y_{Qi}$ are (possibly complex) dimensionless Yukawa couplings, one for each down-type quark, which are parameters of the model.
The $U(1)_X$ charge of $Q_R$ (and hence also $Q_L$) is then fixed by $U(1)_X$-invariance of (\ref{mass terms}), to be $\hat{Q}_Q  = -\hat{Q}_{\Phi} + \hat{Q}_q$, as recorded in (\ref{charges}).
To simplify our analysis, we shall assume the limit where $m_Q \gg |v_{\Phi} Y_{Qi}|$. 

We may package together these terms into 4 by 4 quark mass matrices. 
We shall assume that we have first rotated in quark space to a basis where the SM 3 by 3 Yukawa matrix $(Y_D)_{ij}$ is diagonalised, {\em viz.} $(Y_D)_{ij}\rightarrow \mathrm{diag}(Y_d,Y_s,Y_b)$ for the down quarks. The resulting 4 by 4 mass matrix for the down quarks is then given by \begin{equation}
\mathcal{L} \supset - \overline{d'}^A_L \left( \mathcal{M}_d \right)_{AB}  d'^B_R, \quad 
\text{where}
\quad \mathcal{M}_d =
\left(
\begin{array}{cccc}
m_d & 0 & 0 & \epsilon_d \, m_Q \\
0 & m_s & 0 & \epsilon_s \, m_Q \\
0 & 0 & m_b & \epsilon_b \, m_Q \\
0 & 0 & 0 & m_Q 
\end{array}
\right),
\end{equation}
where $A= \{ 1,2,3,4 \}$, $m_i=\frac{vY_i}{\sqrt{2}}$ (where $v$ is the usual Higgs vev), and the mixing parameters $\epsilon_i \equiv \frac{v_{\Phi} Y_{Qi}}{m_Q \sqrt{2}}\ll 1 $, given our limit of large $m_Q$. We here use notation where the primed fields are current eigenstates, while the unprimed ones denote the physical mass eigenstates. We may diagonalise this 4 by 4 mass matrix completely with a further bi-unitary transformation of the form
\begin{equation}
\begin{cases}
d'_L = U_L d_L \\
d'_R = U_R d_R 
\end{cases}
\quad \rightarrow \quad
U_L^{\dagger} \mathcal{M}_d U_R = \hat{\mathcal{M}}_d  
\quad \rightarrow \quad
\begin{cases}
U_L^{\dagger} \mathcal{M}_d \mathcal{M}^{\dagger}_d U_L = \hat{\mathcal{M}}^2_d,  \\
U_R^{\dagger} \mathcal{M}^{\dagger}_d \mathcal{M}_d U_R = \hat{\mathcal{M}}^2_d,
\end{cases}
\end{equation}
hence the unitary matrices $U_L$ and $U_R$ can be extracted from the eigenvectors of $\mathcal{M}_d \mathcal{M}^{\dagger}_d$ and $\mathcal{M}^{\dagger}_d \mathcal{M}_d$ respectively. To leading order in the small expansion parameters $\epsilon_i$ we find the simple expressions
\begin{equation}
U_L = 
\left(
\begin{array}{cccc}
1 & 0 & 0 &  \epsilon_d \\
0 & 1 & 0 &  \epsilon_s \\
0 & 0 & 1 &  \epsilon_b \\
-\epsilon_d^* & -\epsilon_s^* &  -\epsilon_b^*  & 1 
\end{array}
\right)
+ \mathcal{O} \left(\epsilon^2,\frac{m_i \epsilon_i}{m_Q}\right),
\qquad
U_R = 
\left(
\begin{array}{cccc}
1 & 0 & 0 & 0 \\
0 & 1 & 0 & 0\\
0 & 0 & 1 & 0 \\
0 & 0 & 0 & 1 
\end{array}
\right)
+ \mathcal{O} \left( \frac{m_i \epsilon_i}{m_Q} \right)
\end{equation}
Terms of $\mathcal{O} (\epsilon^2)$ in $U_L$ are in principle important when discussing flavour violation, but as we shall soon see, we do not in fact need to compute them. Concerning the rotation in the right-handed down quarks, we keep $U_R=1$, justified by the smallness of the quark masses compared to $m_Q$.

From these mixing matrices, we can compute, from now on neglecting higher order terms in the $\epsilon_i$ parameters, the couplings of the $Z^\prime$\footnote{In the following formulae, $Z^\prime$ denotes the physical heavy gauge boson, which is a mass eigenstate. As we shall soon see, this is equal to the gauge field for $U(1)_X$, which shall be denoted $X_\mu$, up to small corrections $\sim \mathcal{O}(m_Z^2/m_{Z^\prime}^2)$.
} to down quarks within our model. Because the SM quarks have universal $U(1)_X$ charges, in the primed basis defined above (in which the SM Yukawa matrices $Y_D$ and $Y_U$ are already diagonalised), we have the following hadronic couplings of the $Z^\prime$,
\begin{eqnarray} \label{hadronic Z prime couplings}
\mathcal{L}_{qZ^\prime} = g_X Z^\prime_\alpha \delta_{ij} \left( \hat{Q}_q\ \overline{d'}^i \gamma^{\alpha} P_L d'^j +\hat{Q}_q\ \overline{u'}^i \gamma^{\alpha} P_L u'^j + \hat{Q}_d\ \overline{d'}^i \gamma^{\alpha} P_R d'^j + \hat{Q}_u\ \overline{u'}^i \gamma^{\alpha} P_R u'^j \right) \nonumber \\
+ g_X  Z^\prime_\alpha \hat{Q}_Q \left( \overline{d'}^4 \gamma^{\alpha} P_L d'^4 +\overline{u'}^4 \gamma^{\alpha} P_L u'^4 +  \overline{d'}^4 \gamma^{\alpha} P_R d'^4 + \overline{u'}^4 \gamma^{\alpha} P_R u'^4 \right),
\end{eqnarray}
where $g_X$ denotes the gauge coupling for $U(1)_X$.
Rotating to the mass basis, we obtain the following couplings of the $Z^\prime$ to the three SM quark families, indexed as usual by $i\in\{1,2,3\}$,
\begin{equation}
\mathcal{L}_{qZ^\prime} \supset g_X Z^\prime_\alpha \left( L^{(d)}_{ij} \, \overline{d}^i \gamma^{\alpha} P_L d^j + R^{(d)}_{ij} \, \overline{d}^i \gamma^{\alpha} P_R d^j 
+ L^{(u)}_{ij} \, \overline{u}^i \gamma^{\alpha} P_L u^j + R^{(u)}_{ij} \, \overline{u}^i \gamma^{\alpha} P_R u^j \right),
\end{equation}
where the 3 by 3 matrices of $Z^\prime$ couplings to the down quarks are given by
\begin{equation}
L^{(d)}_{i j} 
= (U_L)^{\dagger}_{iA} 
\left(
\begin{array}{cccc}
\hat{Q}_q & 0 & 0 & 0 \\
0 & \hat{Q}_q & 0 & 0\\
0 & 0 & \hat{Q}_q & 0 \\
0 & 0 & 0 & \hat{Q}_Q
\end{array}
\right)_{AB}
(U_{L})_{Bj}, \qquad R^{(d)}_{i j} = \hat{Q}_d \delta_{ij}.
\end{equation}
The matrices of $Z^\prime$ couplings to the up quarks are
\begin{equation}
L^{(u)}_{i j} = \left(V L^{(d)}_{i j} V^{\dagger} \right)_{ij}, \qquad R^{(u)}_{i j} = \hat{Q}_u \delta_{ij}
\end{equation}
where $V$ is the CKM matrix, and the index $A\in\{1,2,3,4\}$.

To extract the flavour violating effects, which are of order $\mathcal{O} (\epsilon^2)$, we can subtract from the 4 by 4 matrix $\tilde{L}_{AB}^{(d)}\equiv \text{diag}(\hat{Q}_q,\hat{Q}_q,\hat{Q}_q,\hat{Q}_Q)$ (defined such that $L^{(d)}_{i j} =(U_L)^{\dagger}_{iA} \tilde{L}_{AB}^{(d)}(U_{L})_{Bj}$) the flavour universal component equal to $\hat{Q}_q \delta_{AB}$. Using also the unitarity of $U_L$, 
we then extract the flavour violating piece of $L^{(d)}_{i j}$ to be (where $i \neq j$):
\begin{equation}
(U_L)^{\dagger}_{iA} 
\left(
\begin{array}{cccc}
0 & 0 & 0 & 0 \\
0 & 0 & 0 & 0\\
0 & 0 & 0 & 0 \\
0 & 0 & 0 & \hat{Q}_Q - \hat{Q}_q
\end{array}
\right)_{AB}
(U_L)_{Bj} =  (\hat{Q}_Q - \hat{Q}_q) \epsilon_i \epsilon^{*}_j = -\frac{Y_{Q_i} Y^*_{Q_j}}{2 m_Q^2} \, v^2_{\Phi}
\end{equation}
where we used the fact that $\hat{Q}_Q - \hat{Q}_q = -\hat{Q}_{\Phi}=-1$.
Of particular interest from the point of view of the NCBAs, there is a coupling of the $Z^\prime$ to $b\bar{s}$, of the form
\begin{equation}
\mathcal{L}_{qZ^\prime} \supset g_{sb} \bar{s}_L Z^\prime_\alpha \gamma^\alpha b_L, \quad \text{where} \quad g_{sb}= -g_X v^2_{\Phi} \frac{Y_{Q_b} Y^*_{Q_s}}{2 m_Q^2}.
\end{equation}
Of course, within our setup there is {\em no} rotation between the mass and weak eigenbasis for the charged leptons because the $U(1)_X$-invariant Yukawa terms are strictly diagonal. Thus, we only have diagonal couplings of the $Z^\prime$ to charged leptons, with the charges as defined in (\ref{charges}), which result in LFUV without LFV.

\subsection{\texorpdfstring{$Z-Z^\prime$}{Z-Z'} mixing}\label{ZZ' mixing}

Provided $a_Y\neq 0$, the Higgs carries $U(1)_X$ charge in our framework of models. This leads to mass mixing between the SM $Z$ boson and the $Z^\prime$, which shall ultimately result in the $Z$ inheriting some small flavour non-universality in its couplings to leptons. The universality of $Z$ couplings to leptons is tightly constrained by precision measurements at LEP, which shall provide an important bound on our model when $a_Y\neq 0$. To derive the following formulae for this $Z-Z^\prime$ mixing, we closely follow Refs.~\cite{Allanach:2018lvl,Allanach:2019iiy}.

As we have already set out, the $U(1)_X$ gauge symmetry is spontaneously broken by the SM singlet complex scalar $\Phi$ acquiring its vev. We shall here denote the original $U(1)_X$ gauge boson by $X_\mu$, reserving the name $Z^{\prime}_\mu$ for the physical boson which is a mass eigenstate.
The mass terms for the heavy gauge bosons come, of course, from the kinetic terms of the scalar fields $H$ and $\Phi$,
\begin{equation}
\mathcal{L}_{H\Phi~\text{kin}} = (D^\mu H)^{\dagger}(D_\mu H) + (D^\mu \Phi)^{*}(D_\mu \Phi), \label{scalarKineticTerms}
\end{equation}
where the covariant derivatives are
\begin{equation}
D_{\mu}H=\partial_\mu H-i\left(\frac{g}{2}\tau^a W^a_{\mu}-\frac{g'}{2}B_\mu - \frac{a_Y g_X}{2} X_\mu\right)H, \qquad D_{\mu}\Phi=(\partial_\mu -i\hat{Q}_{\Phi} g_X X_\mu)\Phi.
\end{equation}
Here, as usual, $g$ and $g'$ denote the gauge couplings for $SU(2)_L$ and $U(1)_Y$ respectively.
Expanding the scalar fields about their vevs in (\ref{scalarKineticTerms}), we find the following mass matrix for the neutral gauge bosons:
\begin{equation} \label{neutral masses}
\mathcal{M}^2_N=\frac{v^2}{4}\left( \begin{array}{ccc}
g'^2 & -gg' & a_Yg'g_X \\
-gg' & g^2 & -a_Ygg_X \\
a_Yg'g_X & -a_Ygg_X & a_Y^2g_X^2\left(1+4\hat{Q}_\Phi^2 v_\Phi^2/a_Y^2v^2\right) \\
\end{array}\right).
\end{equation}
We rotate to the mass basis of physical neutral gauge bosons, $(A_\mu,Z_\mu,Z^{\prime}_\mu)^T\equiv{\bf A_\mu}=O^T {\bf A'_\mu}$, where the orthogonal matrix $O$ is
\begin{equation}
O=
\left( \begin{array}{ccc}
\cos\theta_w & -\sin\theta_w \cos\alpha_z & \sin\theta_w \sin\alpha_z \\
\sin\theta_w & \cos\theta_w \cos\alpha_z & -\cos\theta_w \sin\alpha_z \\
0 & \sin\alpha_z & \cos\alpha_z \\
\end{array}\right), \label{orthogonal}
\end{equation}
where $\theta_w$ is the Weinberg angle (such that $\tan\theta_w=g'/g$), and $\alpha_z$ is the $Z$-$Z^\prime$ mixing angle. The masses of the $Z$ and $Z'$ boson are then the two non-zero eigenvalues of the above mass matrix. The mass of the $Z'$ is
\begin{equation}
m_{Z^\prime}\approx \frac{va_Yg_X}{2}\sqrt{1+\frac{4\hat{Q}_\Phi^2 v_\Phi^2}{a_Y^2 v^2}} \approx g_X|\hat{Q}_\Phi| v_\Phi=g_X v_\Phi, \label{mass}
\end{equation}
and the mixing angle is
\begin{equation}
\sin\alpha_z \approx \frac{a_Yg_X}{\sqrt{g^2+g'^2}}\left(\frac{m_Z}{m_{Z^\prime}}\right)^2, \label{mixing}
\end{equation}
where we are working in the limit that $m_{Z'} \gg m_Z$. 

\bigskip

\begin{figure}
\centering
\includegraphics[width=0.47 \textwidth]{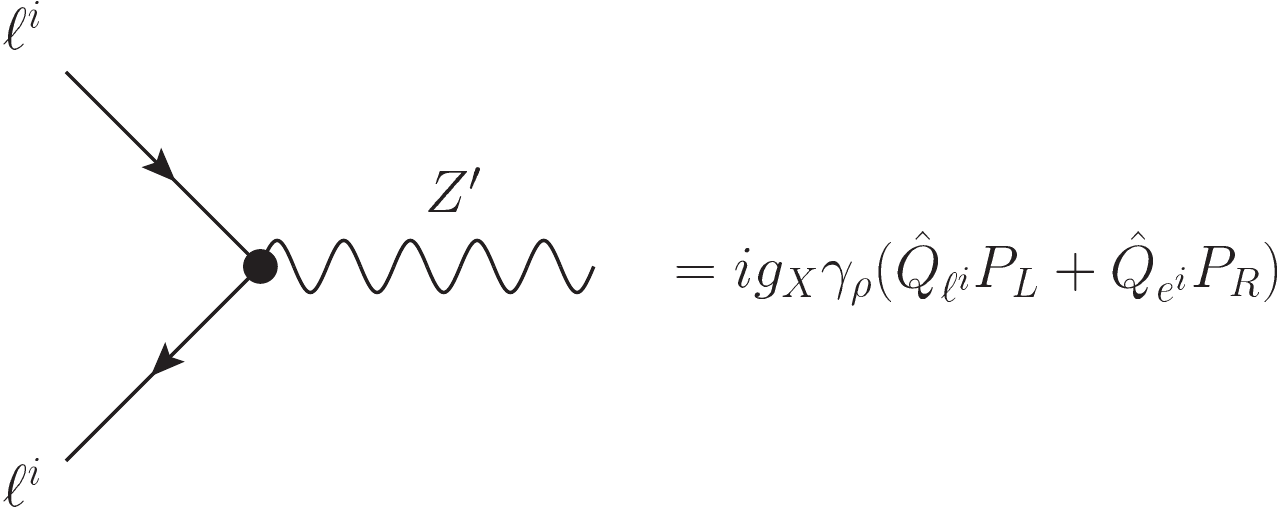} ~~~~~~
\includegraphics[width=0.47 \textwidth]{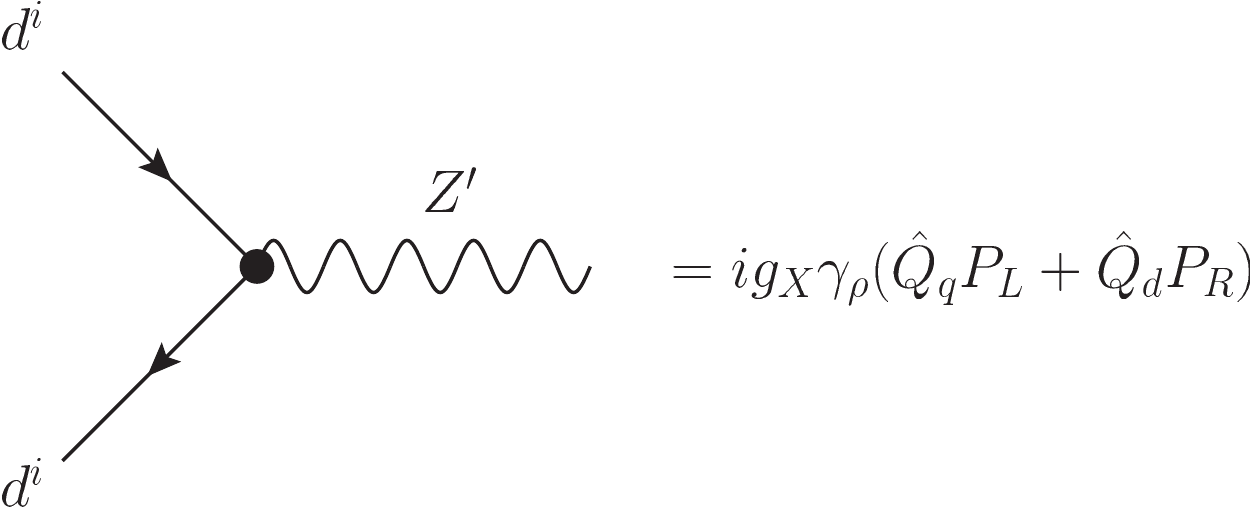} \\[12pt]
\includegraphics[width=0.47 \textwidth]{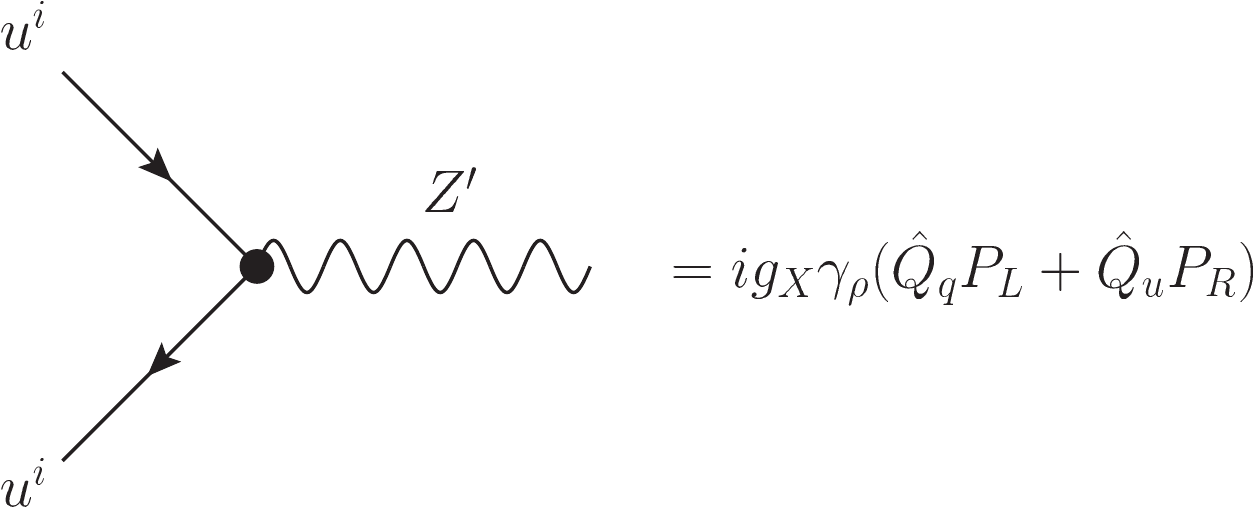} ~~~~~~
\includegraphics[width=0.33 \textwidth]{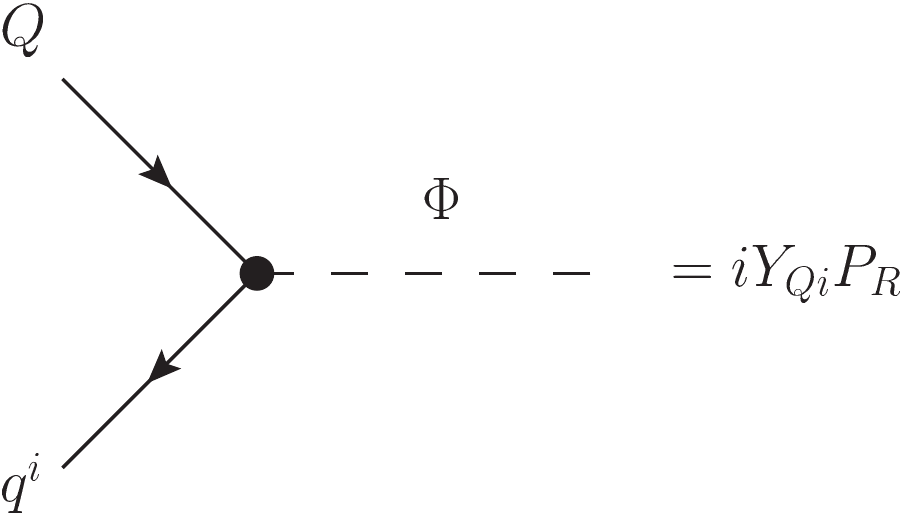} ~~~~~~~~~~~~~~~~~~~\\[12pt]
\includegraphics[width=0.37 \textwidth]{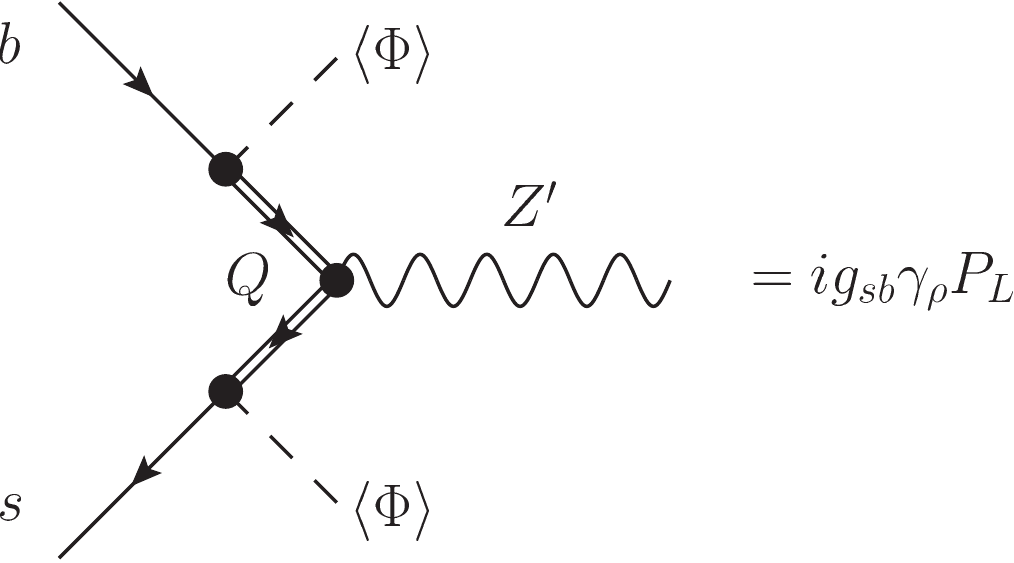}
\caption{The Feynman rules associated with the most important couplings of the $Z^\prime$ and the scalar SM singlet field $\Phi$ to the SM quarks and leptons. Top left: coupling of the $Z^\prime$ to leptons; top right/center left: leading flavour diagonal couplings of the $Z^\prime$ to quarks; center right: couplings of $\Phi$ to the quarks and the vector-like fermions; bottom: effective flavour violating couplings of the $Z^\prime$ to quarks.  
\label{fig:diagrams}}
\end{figure}

This concludes the description of our framework of models. In summary, the full lagrangian may be written as
\begin{equation} \label{full lagrangian}
\mathcal{L} = \mathcal{L}_{\text{SM}} -\frac{1}{4}X_{\mu\nu}X^{\mu\nu} + \mathcal{L}_{\text{mix}}+ \mathcal{L}_{qZ^\prime} + \mathcal{L}_{\ell Z^\prime}+ \mathcal{L}_{H\Phi~\text{kin}}-V(H,\Phi)+\mathcal{L}_{\text{dark}},
\end{equation}
where $\mathcal{L}_{\text{SM}}$ denotes the SM lagrangian (but with the Higgs kinetic and potential terms removed), $X_{\mu\nu}=\partial_\mu X_\nu - \partial_\nu X_\mu$,\footnote{Note that we have assumed that any kinetic mixing between the $Z$ and $Z^\prime$ gauge fields is set to zero.} $\mathcal{L}_{\text{mix}}$ may be read off from (\ref{mass terms}), $\mathcal{L}_{qZ^\prime}$ from (\ref{hadronic Z prime couplings}), $\mathcal{L}_{H\Phi~\text{kin}}$ from (\ref{scalarKineticTerms}), and $\mathcal{L}_{\ell Z^\prime}$ denotes the (flavour-diagonal, but flavour non-universal) $Z^\prime$ coupling to leptons, given by
\begin{equation}
\mathcal{L}_{\ell Z^\prime} = g_X Z^\prime_\alpha \sum_{i=1}^3 \left( \hat{Q}_{\ell^i}\ \overline{\ell}^i \gamma^{\alpha} P_L \ell^i +\hat{Q}_{\ell^i}\ \overline{\nu}^i \gamma^{\alpha} P_L \nu^i + \hat{Q}_{e^i}\ \overline{\ell}^i \gamma^{\alpha} P_R \ell^j + \hat{Q}_{\nu^i}\ \overline{\nu}^i \gamma^{\alpha} P_R \nu^j \right),
\end{equation}
where we have included here the $Z^\prime$ couplings to the dark states $\nu^i_R$.
We summarise the Feynman rules associated with the important new physics couplings of the $Z^\prime$ boson to SM fermions in Fig.~\ref{fig:diagrams}, for the reader's convenience.
Finally, $V(H,\Phi)$ is a scalar potential which determines the vevs of $H$ and $\Phi$ (but which we do not specify beyond that), and $\mathcal{L}_{\text{dark}}$ denotes any terms in the lagrangian involving dark states only.
All that remains to specify the model fully is a discussion of how the neutrino mass sector may arise. Such a discussion involves a rather in-depth analysis of the dark sector of the theory (encoded in $\mathcal{L}_{\text{dark}}$), and would be something of a digression at this stage, so we relegate this discussion to Appendix \ref{neutrinos}. 

The following two Sections are devoted to exploring the phenomenological consequences of this setup.

%%%%%%%%%%%%%%%%%%%%%%%%%%%%%%%%%
%%%%%%%%%%%%%%%%%%%%%%%%%%%%%%%%%%

\section{Implications for the \texorpdfstring{$B$}{B} anomalies} \label{anomalies}

After integrating out the heavy $Z^\prime$ boson, we obtain BSM contributions to a host of dimension six Wilson operators in the SMEFT which are capable of explaining the observed NCBAs, depending on the values of the coefficients $a_e$, $a_\mu$, $a_\tau$, and $a_Y$. In \S \ref{B anatomy} we will present the contributions to the relevant Wilson coefficients within our framework of models, which we shall see carry a particular structure allowing both vectorial and axial currents (but where the axial contributions must be lepton flavour universal). We then present results of our global fits to the NCBA data using \texttt{flavio}~\cite{Straub:2018kue} in \S \ref{fits}, after making a number of (physically well-motivated) simplifying assumptions to cut down our parameter space.

\subsection{The anatomy of the \texorpdfstring{$B$}{B} anomalies within our framework} \label{B anatomy}

At a generic point in the parameter space we are considering, the following four operators (all of which couple only to the left-handed $b\bar{s}$ current) are all present:
\begin{eqnarray}
  {\mathcal L}_{bs\ell\ell} =
  C_L^e(\overline{s_L} \gamma_\rho b_L) 
  (\overline{e_L} \gamma^\rho e_L) +
  C_R^e(\overline{s_L} \gamma_\rho b_L) 
  (\overline{e_R} \gamma^\rho e_R) +\nonumber \\
  C_L^\mu (\overline{s_L} \gamma_\rho b_L) 
  (\overline{\mu_L} \gamma^\rho \mu_L) +
  C_R^\mu (\overline{s_L} \gamma_\rho b_L) 
  (\overline{\mu_R} \gamma^\rho \mu_R), \label{bsll2}
\end{eqnarray}
where the coefficients are
\begin{eqnarray}
C^{\alpha}_L &=& -\frac{Y_{Qb} Y^*_{Qs}}{2 m_Q^2}\hat{Q}_{\ell^{\alpha}} = -\frac{Y_{Qb} Y^*_{Qs}}{2 m_Q^2} \, \left( a_{\alpha} - a_Y/2 \right)  \\ 
C^{\alpha}_R &=& -\frac{Y_{Qb} Y^*_{Qs}}{2 m_Q^2}\hat{Q}_{e^{\alpha}}    = -\frac{Y_{Qb} Y^*_{Qs}}{2 m_Q^2} \, \left( a_{\alpha} - a_Y \right) 
\end{eqnarray}
where $\alpha$ is the lepton flavour index $\alpha= \{ e,\mu,\tau \}$. We may convert these Wilson coefficients into the more conventional basis of vectorial and axial currents, corresponding to $C^\alpha_9 \equiv (C^\alpha_L+C^\alpha_R)/2$ and $C^\alpha_{10} \equiv -(C^\alpha_L-C^\alpha_R)/2$, for which we find:
\begin{eqnarray}
C^{\alpha}_9 &=&  -\frac{Y_{Qb} Y^*_{Qs}}{2 m_Q^2} \, \left( a_{\alpha} - \frac{3}{4} a_Y \right), \label{C9} \\
C^{\alpha}_{10} &=&  \frac{Y_{Qb} Y^*_{Qs}}{8 m_Q^2} \, a_Y. \label{C10}
\end{eqnarray}
This choice of basis makes clear that the NCBAs have an interesting, and simple, structure within our framework of models. A few noteworthy points are:
\begin{itemize}
\item The LFUV must come entirely from the vectorial current.
\item There may nonetheless be an axial current contribution, but this is lepton flavour universal. Indeed, this feature can be traced back to our initial assumptions regarding the renormalisable Yukawa sector, which implied that the differences between the left-handed and right-handed lepton charges were family universal.
\item The presence of this axial contribution requires $a_Y\neq 0$, which then implies (i) that there is $Z$-$Z^\prime$ mixing (since $\hat{Q}_H =-a_Y/2$), and so important constraints from LEP lepton flavour universality measurements, and (ii) that there are necessarily couplings of the $Z^\prime$ to valence quarks, as can be deduced from the charges in (\ref{charges}), and so important constraints from LHC direct searches in say $pp\rightarrow \mu\mu$.
\item If we wish to remove couplings of the $Z^\prime$ to quarks in order to loosen the constraints from direct searches, we require both that $a_Y=0$ (thus removing any axial component in the Wilson coefficients, as above) {\em and} that $a_e+a_\mu+a_\tau=0$. Thus, in this limit, there are only two independent parameters, which we can choose to be any two of $(a_e,a_\mu,a_\tau)$.
\item The expressions (\ref{C9}, \ref{C10}) for the Wilson coefficients do not depend on the parameter $v_\Phi$, but only on the parameters from the gauge sector and from the mixing sector. The dependence on both $v_\Phi$ and $g_X$ happens to cancel between the factors of $M_{Z^\prime}$ in the denominator and the couplings in the numerator.
\end{itemize}

\subsection{Global fits at a benchmark point in the parameter space} \label{fits}

Within this general class of models we have formulated, there is a large number of {\em a priori} free parameters. We have:
\begin{itemize}
\item From the gauge sector: $g_X, a_e, a_{\mu}, a_{\tau}, a_Y.$ 
\item From the mixing sector: $Y_{Qd},Y_{Qs},Y_{Qb}, m_Q.$
\item From the scalar sector: $v_{\Phi} + \dots,$ 
\end{itemize}
where the $\dots$ indicates additional parameters appearing the extended scalar potential, which we shall not be concerned with here.
To thoroughly explore the phenomenology in all these parameters is a complicated task; in this paper we shall not attempt such a complete phenomenological characterisation of this parameter space. Rather, we prefer to make a number of well-motivated simplifying assumptions to cut down the parameter space. In this way, we shall define a ``benchmark'' region in our parameter space which is most relevant for the $B$ anomalies and related phenomenology, and we shall find that there is room  in this region to evade all current experimental constraints while remaining highly predictive.

The assumptions we make are as follows:
\begin{enumerate}
\item $a_{\mu}=1$.
This is in some sense a choice of normalisation, also made in Ref.~\cite{Altmannshofer:2014cfa}, which forces a non-zero new physics contribution in the coupling of the $Z^\prime$ to muons. While there remains a logical possibility for fitting the `theoretically clean' ratios $R_{K^{(\ast)}}$ with $a_\mu=0$ and new physics only in the electron couplings, the inclusion of other NCBA data (for example $P_5^{\prime}$ in $B\rightarrow K^\ast \mu\mu$ decays, and $BR(B_s\rightarrow \mu\mu)$) strongly favours new physics  (or at least a sizable component) in the muon channel.

\item $a_{\tau}=0$.
The NCBAs concern only electrons and muons, and so are largely insensitive to the tauon couplings. Thus, the value of $a_\tau$ shall have no effect on the global fits we will perform, and so it is convenient to set $a_{\tau}=0$ at the outset.

\item $Y_{Qd}=0$.
Fitting the NCBAs requires both $Y_{Qs}$ and $Y_{Qb}$ are non-zero, but does not require a non-zero coupling $Y_{Qd}$. By choosing to set $Y_{Qd}=0$, we prevent new flavour violation beyond the SM in processes involving the down quark. With this choice we are therefore automatically safe from the bounds on {\em e.g.} kaon and $B_d$ mixing. Note that this assumption, like the others we are making, is not forced upon us by the data, but is a sensible choice which we make to reduce the parameter space.

\item $g_X$ is irrelevant for low energy processes.
In the limit $|Y_{Qi} v_{\Phi}| < m_Q$ which we are assuming, all the effects induced by the $Z^\prime$ exchange are independent of $g_X$ (since two powers of $g_X$ from the gauge vertices in the numerator are always cancelled by two powers of $g_X$ in the denominator from the mass of the gauge boson $m_{Z^\prime}=g_X v_\Phi$). If the mass of the $Z^\prime$ is not too high, then the coupling $g_X$ and $v_\Phi$ do indeed become two independent parameters, since bump searches for the $Z^\prime$ depend on $m_{Z^\prime}$ itself. Nonetheless, if we are in the r\'egime $m_{Z'} \gg m_{Z}$,\footnote{More precisely, if the mass of the $Z^\prime$ is greater than that probed by direct searches, which is currently $\sim$ 6 TeV~\cite{Aad:2019fac}. We discuss these direct searches in \S \ref{direct search}.} then direct searches for the $Z^\prime$ only depend on the contact 4-fermion operators, independent of $g_X$. We shall see in \S \ref{direct search} that there is such a r\'egime in our parameter space in which the $Z^\prime$ couplings remain perturbative (in the sense that the $Z^\prime$ width is less than, say, 30\% of its mass, or thereabouts).

\item $Y_{Qb} Y^*_{Qs}$ as a single parameter.
Each of the couplings $Y_{Qb}$ and $Y_{Qs}$ appear only in the combination $Y_{Qb} Y^*_{Qs}$ in both the Wilson coefficients $C_9^\alpha$ and $C_{10}^\alpha$ given in (\ref{C9}, \ref{C10}), and in the BSM contribution to $B_s$ mixing (see \S \ref{Bs mixing}), which provides the main constraint sensitive to the quark flavour violation.

\end{enumerate}

\begin{figure}
\centering
\includegraphics[width=1.0 \textwidth]{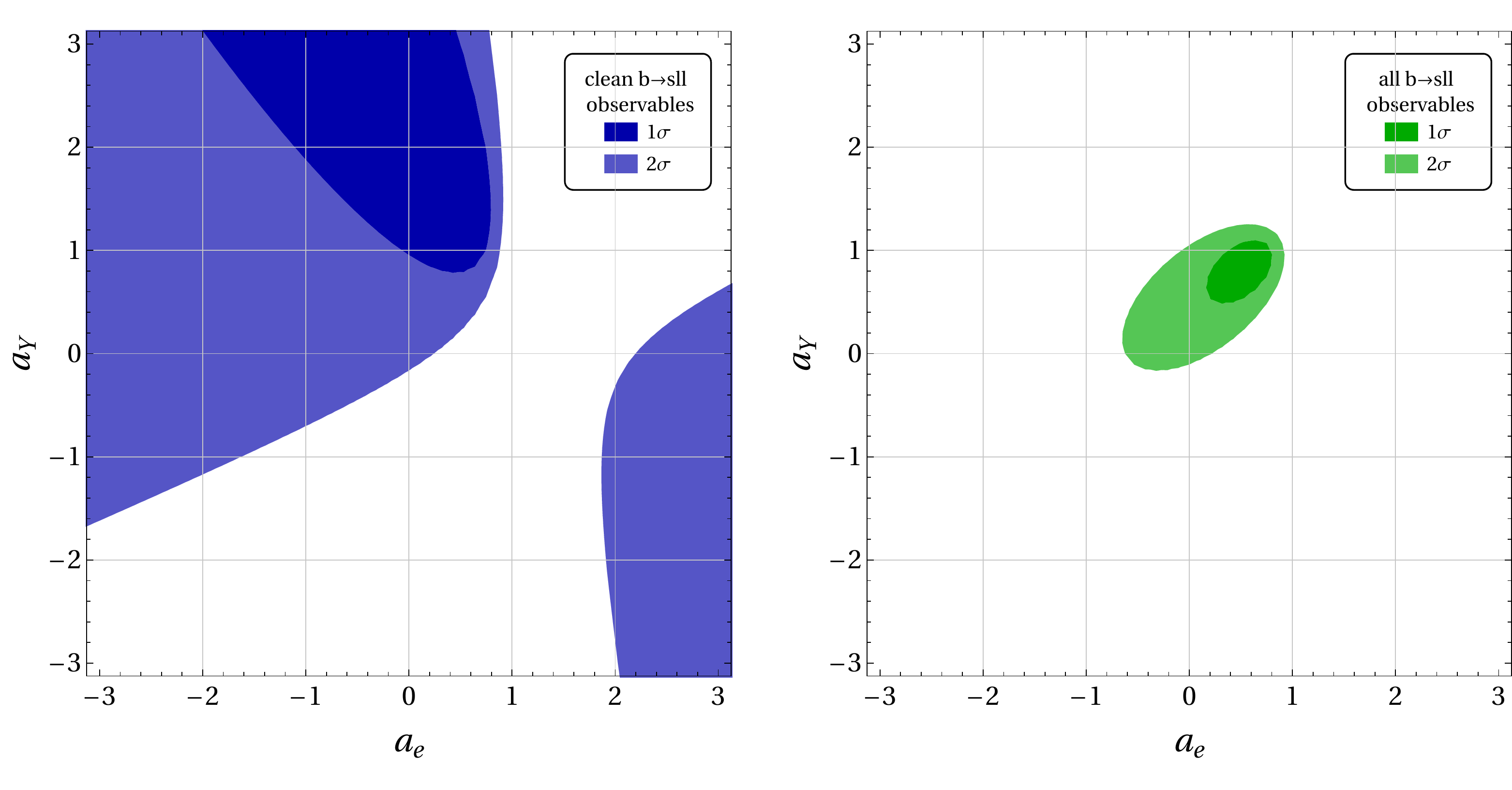}
\caption{Global fits showing the 1$\sigma$ and 2$\sigma$ regions in the $a_e$ {\em vs.} $a_Y$ plane, for fixed $a_\mu=1$ and $a_\tau=0$, from a Gaussian approximation of the likelihood that was generated using \texttt{flavio}. For the plot on the left-hand-side, only the observables $R_{K}$, $R_{K^{\ast}}$, $BR(B_s\rightarrow \mu\mu)$, and $BR(B \rightarrow X_s \ell\ell)$ were included in the fit. For the plot on the right-hand-side, we also include all the branching ratios and $CP$-averaged angular observables of exclusive semileptonic $b$ decays into the fit.
\label{fig:fits}}
\end{figure}

Given these assumptions, we perform global fits to the NCBA data in the plane of $a_e$ versus $a_Y$, using \texttt{flavio}~\cite{Straub:2018kue}. The remaining physical parameters affect the NCBAs only through the overall normalisation of the Wilson coefficients (via the combination $Y_{Qb} Y^*_{Qs}/ m_Q^2$) which, for the purpose of fitting, we write as
\begin{eqnarray} \label{eq:C1}
C^{\alpha}_9 &=&  \frac{4 G_F}{\sqrt{2}} V_{tb} V_{ts}^* \frac{\alpha_\text{em}}{4 \pi} C \, \left( a_{\alpha} - \frac{3}{4} a_Y \right), \\ \label{eq:C2}
C^{\alpha}_{10} &=&  - \frac{4 G_F}{\sqrt{2}} V_{tb} V_{ts}^* \frac{\alpha_\text{em}}{4 \pi} C \, \frac{a_Y}{4} .
\end{eqnarray}
The normalisation $C$ is extracted point-by-point from the fit. 

The results of the fit are shown in Fig.~\ref{fig:fits}. Details of the fit procedure are given in Appendix~\ref{app:fit}. In the plot on the left-hand-side, only a subset of `clean observables' ({\em i.e.} for which the theoretical uncertainties are smaller) are included in the fit. These clean observables are $R_{K}$, $R_{K^{\ast}}$, and the branching ratios $BR(B_s\rightarrow \mu\mu)$, $BR(B \to X_s \mu\mu)$ and $BR(B \to X_s ee)$. The dark (light) blue region extends to the 1$\sigma$ (2$\sigma$) best fit contour. Interestingly, we find that a very large region of the parameter space in $(a_e,a_Y)$ allows a good fit to these clean LFU ratios. Indeed, we find two disconnected `lobes' in the parameter space which fit the data at the 2$\sigma$ level. In the plot on the right-hand-side, we further include all the branching ratios of the exclusive semileptonic branching ratios $B \to K \mu\mu$, $B \to K^*\mu\mu$, $B_s \to \phi \mu\mu$ and $\Lambda_b \to \Lambda \mu\mu$, as well as all $CP$-averaged angular observables in those decay modes into the fit. We see that the inclusion of additional variables beyond  $R_{K^{(\ast)}}$ has a dramatic effect on the fit, as indicated by the shaded 1$\sigma$ and 2$\sigma$ regions in green, where now a definite quasi-elliptical region in parameter space is selected. From hereon, we shall choose this best-fit point as our primary `benchmark' for further study, at which we shall consider the impact of other important experimental constraints in \S \ref{pheno}. The best-fit values for $a_e$, $a_Y$, and the overall normalisation $C$ are recorded in Table~\ref{tab:bestfit}.

\def\arraystretch{1.5}
\setlength\tabcolsep{8pt}
\begin{table}
  \begin{center}
    \begin{tabular}{ cccccc}    
      \hline\hline
      Observables in fit & $a_e$ & $a_\mu$ & $a_\tau$ & $a_Y$ & $C$ \\ 
      \hline
      Clean & 0.49 & 1 & 0 & 1.66 & 1.51 \\
      \hline
      All & 0.59 & 1 & 0 & 0.87 & 2.01 \\
      \hline\hline
    \end{tabular}
  \end{center} \caption{\label{tab:bestfit} Best-fit values for $a_e$ and $a_Y$, subject to the choices $a_\mu=1$ and $a_\tau=0$. The value $C$ gives the overall normalisation of the Wilson coefficients at the best-fit point, see eqs.~(\ref{eq:C1}) and (\ref{eq:C2}).
The first line is for the fit to a subset of `clean observables', corresponding to the left-hand-plot of Fig.~\ref{fig:fits}. The second line, for all observables, corresponds to the right-hand-plot of Fig.~\ref{fig:fits}. We select the latter best-fit point as our primary `benchmark point' in parameter space from hereon.
}
\end{table}

\def\arraystretch{1.5}
\setlength\tabcolsep{8pt}
\begin{table}
  \begin{center}
     \begin{tabular}{cccccccccc}
    \hline\hline
      $\hat{Q}_q$ & $\hat{Q}_u$ & $\hat{Q}_d$ & $\hat{Q}_{\ell^1}$ & $\hat{Q}_{\ell^2}$ & $\hat{Q}_{\ell^3}$ & $\hat{Q}_{e^1}$ & $\hat{Q}_{e^2}$ & $\hat{Q}_{e^3}$ & $\hat{Q}_H$ \\ 
      \hline
      -0.03 & 0.40 & -0.47 & 0.16 & 0.57 & -0.44 & -0.28 & 0.13 & -0.87 & -0.44  \\
      \hline\hline
    \end{tabular}
  \end{center} \caption{\label{tab:charges} Charges for all the SM fields at our benchmark point in parameter space $(a_e,a_\mu,a_\tau,a_Y)=(0.59,1,0,0.87)$, which we obtained by fitting for $(a_e,a_Y)$ subject to $a_\mu=1$, $a_\tau=0$, and using all observables in the fit. This corresponds to the bottom line in Table~\ref{tab:bestfit}.
}
\end{table}

We can draw some interesting conclusions from these global fits, particularly from the fit to all observables which appears to select a clear preferred region. Firstly, while we see that a reasonable fit can be obtained with new physics only in the muon (which corresponds to the point $a_e=a_Y=0$, which lies within the 2$\sigma$ contour), there is strong pull to include some new physics component in the electron. Furthermore, independently of $a_e$, we see that the fit also favours turning on a significant component $a_Y>0$, which we know gives a flavour-universal contribution to $C_9$ and $C_{10}$.

\section{Other phenomenological constraints} \label{pheno}

In addition to fitting the NCBAs, there are several important phenomenological constraints on our model. These come from high-$p_T$ LHC searches for the $Z^\prime$ ({\em e.g.} in $pp\rightarrow \mu\mu$), $B_s$ meson mixing, a number of electroweak precision observables (in particular, LFU precision measurements in the $Z$ boson couplings from LEP, and the measurement of the $\rho$-parameter), and neutrino trident production. In this Section we shall consider each of these constraints in turn.

In the same spirit as above, our goal here is not to characterise the phenomenology of these models in all detail, but rather to analyse the general structure of the different constraints within our framework, and to show as a `proof of principle' that there is a viable parameter space. Thus, we shall present the theoretical expressions relevant to each constraint and comment on their form, and we are content to extract the numerical bounds only at our benchmark point in the parameter space. Recall that this benchmark point corresponds to the second line of Table~\ref{tab:bestfit}, for which the charges are listed in Table~\ref{tab:charges}.

\subsection{Direct searches at the LHC} \label{direct search}

The strongest $Z^\prime$ bounds from direct searches at the LHC can be obtained by interpreting recent ATLAS searches for resonances in both the $pp\rightarrow \mu^+ \mu^-$ and $pp\rightarrow e^+ e^-$ channels
in 139 fb$^{-1}$ of 13 TeV data~\cite{Aad:2019fac}, which extend up to a dilepton invariant mass of 6 TeV.\footnote{There are also less constraining $Z^\prime$ searches from ATLAS on smaller data sets in other channels, such as a pair of $Z^\prime\rightarrow t\bar{t}$ searches in 36.1 fb$^{-1}$ of 13 TeV data~\cite{Aaboud:2018mjh,Aaboud:2019roo} which extend out to 5 TeV, and a $Z^\prime\rightarrow \tau^+\tau^-$ search in 10 fb$^{-1}$ of 8 TeV data~\cite{Aad:2015osa} which extends out to 2.5 TeV.  
} The constraints on a number of simple $Z^\prime$ models coming from these ATLAS searches have recently been computed in Ref.~\cite{Allanach:2019mfl}, and if we wished to calculate an accurate bound from direct searches valid for any value of the coupling $g_X$, then we should follow a similar methodology.

For our purposes in this paper, we first note that if $m_{Z^\prime}\approx g_X v_\Phi$ exceeds 6 TeV, then the constraints from these direct bump searches do not apply directly. In this region of parameter space, which we shall refer to as the `contact r\'egime', the high-$p_T$ tail of the dimuon invariant mass-squared distribution is of course still sensitive to the $Z^\prime$, but its effect may be computed using a simple EFT calculation involving the four-fermion effective operators which couple the final state lepton pair to a quark pair in the initial state~\cite{Greljo:2017vvb}. We must first establish that this r\'egime is even accessible within our framework; in other words, we must show that the $Z^\prime$ mass can be as heavy as 6 TeV. 

As we shall see in \S \ref{Bs mixing}, there is an {\em upper} bound on the vev $v_\Phi$ from $B_s$ mixing of approximately $v_\Phi \lesssim 4$ TeV;  there is also an upper bound on the coupling $g_X$, beyond which the $Z^\prime$ becomes strongly coupled and perturbativity breaks down, which we shall now estimate. This bound is approached when the width $\Gamma_{Z^\prime}$ of the $Z^\prime$ resonance becomes broad, which we take to be when $\Gamma_{Z^\prime}/m_{Z^\prime}\approx 0.3$ or so. To calculate $\Gamma_{Z^\prime}$, we need the partial width of the $Z^\prime$ decaying into a pair of massless fermions $f \bar{f}$, which in the limit where $m_{Z^\prime} \gg 2m_t$ is given by $\Gamma_{Z^\prime}^{f \bar{f}}=C g_X^2 \hat{Q}_f^2 m_{Z^\prime}/(24 \pi)$, where $C=3$ for
quarks and $C=1$ for leptons, and $\hat{Q}_f$ denotes the $U(1)_X$ charge of the fermion $f$, as recorded in (\ref{charges}), and, at our benchmark point in parameter space, in Table~\ref{tab:charges}. Summing over all the SM fermion species,\footnote{Other decay modes of the $Z^\prime$, for example to $ZH$, can be shown to be subleading.} we obtain
\begin{equation}
\frac{\Gamma_{Z^\prime}}{m_{Z^\prime}} \approx 0.071 g_X^2, 
\end{equation}
and so perturbativity breaks down for couplings as large as $g_X \sim \sqrt{0.3/0.071} \approx 2.1$. We may therefore consider large-ish couplings in the window $1.5 \lesssim g_X \lesssim 2.1$ for which the contact r\'egime is accessible, with the $Z^\prime$ couplings remaining perturbative. In this r\'egime we can compute the bound on $v_\Phi$ using an EFT approach. This is in some sense the weakest possible bound on $v_\Phi$ coming from direct searches, at least at our benchmark point in parameter space.

In Ref.~\cite{Greljo:2017vvb}, bounds are tabulated for each semi-leptonic four-fermion operator using such an EFT approach, considered one operator at a time, which are valid in this contact r\'egime.\footnote{Note that the bounds in Ref.~\cite{Greljo:2017vvb} were computed using an ATLAS search on 36 fb$^{-1}$ of 13 TeV data from 2017~\cite{Aaboud:2017buh}, which was superseded by the 2019 search published in Ref.~\cite{Aad:2019fac}.} For the benchmark point we are considering however, there are of course multiple relevant four-fermion operators turned on, 
with the dominant couplings due to the following three operators (we adopt the normalisation of Wilson coefficients used in Ref.~\cite{Greljo:2017vvb} for ease of comparison):
\begin{equation}
\frac{C^{(1)}_{q_{11} \ell_{22}}}{v^2}(\bar{q}^1_L\gamma_\mu q^1_L)(\bar{\ell}^2_L \gamma^\mu \ell^2_L), \qquad \frac{C_{d_{11} \ell_{22}}}{v^2}(\bar{d}^1_R\gamma_\mu d^1_R)(\bar{\ell}^2_L \gamma^\mu \ell^2_L), \qquad \frac{C_{u_{11} \ell_{22}}}{v^2}(\bar{u}^1_R\gamma_\mu u^1_R)(\bar{\ell}^2_L \gamma^\mu \ell^2_L).
\end{equation}
Within our framework, the Wilson coefficients are given by 
\begin{equation}
C^{(1)}_{q_{11} \ell_{22}}=\frac{v^2}{2v_\Phi^2}\hat{Q}_q \hat{Q}_{\ell^2}, \qquad C_{d_{11} \ell_{22}}=\frac{v^2}{2v_\Phi^2}\hat{Q}_d \hat{Q}_{\ell^2}, \qquad C_{u_{11} \ell_{22}}=\frac{v^2}{2v_\Phi^2}\hat{Q}_u \hat{Q}_{\ell^2}.
\end{equation}
Using the most recent 139 fb$^{-1}$ ATLAS search described above~\cite{Aad:2019fac}, and including the contributions to the $pp\rightarrow \ell^+\ell^-$ cross-section from all four-fermion operators, one obtains the bound 
\begin{equation}
v_\Phi > 3.1 ~\text{TeV}
\end{equation}
at the 95\% C.L.\footnote{We are very grateful to David Marzocca for performing this calculation and providing us with the result.} We shall see that this provides one of the most important bounds on the model at the benchmark point, alongside bounds from the $\rho$-parameter and from $B_s$ mixing which we compute next.

\subsection{Neutral meson mixing} \label{Bs mixing}

The quark flavour violation that mediates $b\rightarrow s$ transitions, which is a necessary ingredient for explaining the NCBAs, immediately results in a BSM contribution to $B_s$ meson mixing. Within our framework of $Z^\prime$ models there is a contribution from tree level exchange of the $Z^\prime$, in addition to a contribution from scalar box diagrams involving $\Phi$. This is exactly analogous to the meson mixing constraints derived in Ref.~\cite{Altmannshofer:2014cfa}, which we here follow.\footnote{Given our simplifying assumption $Y_{Qd}=0$, there are no analogous bounds from kaon or $B_d$ mixing. There are nonetheless bounds from $D$ mixing, though these are less constraining than those from $B_s$ mixing, since the coefficient of the latter is inextricably tied to fitting the NCBAs. In particular, we find that $D$ mixing constraints are much weaker than the ones from $B_s$ mixing as long as there is a slight hierarchy in the couplings, {\em viz.} $|Y_{Qs}| < |Y_{Qb}|$.}

The mixing amplitude $M_{12}$ for the $B_s$ meson system takes the form
\begin{equation}
\frac{M_{12}}{M_{12}^{\text{SM}}} = 1 + \left(\frac{g^4}{16\pi^2 m_W^2}( V_{tb} V_{ts}^\ast)^2 S_0 \right)^{-1} C^{bs}_{LL},
\end{equation}
where $m_W$ is the mass of the $W$ boson and $S_0\simeq 2.3$ is a SM loop function. The Wilson coefficient $C^{bs}_{LL}$ is given by
\begin{equation} \label{BS mixing}
C^{bs}_{LL} = (Y_{Qb} Y^*_{Qs})^2 \left( \frac{v^2_{\Phi}}{m_Q^4} + \frac{1}{16 \pi^2 m_Q^2} \right).
\end{equation}
Here, the first term on the right-hand-side is due to the $Z^\prime$ exchange, which scales somewhat unusually like $v_\Phi^2$; the two powers of $v_\Phi$ in the denominator from the $Z^\prime$ propagator are compensated by four powers of $v_\Phi$ in the numerator arising from the square of the coupling $g_{sb}=-g_X v_\Phi^2 Y_{Qb} Y^*_{Qs} / (2m_Q^2)$.
The second term on the right-hand-side is due to the 1-loop box diagram. 

While the mass difference $\Delta M_s \propto |M_{12}|$ is measured with excellent precision, $\Delta M_s^\text{exp} = (17.757 \pm 0.021) /$ps~\cite{Amhis:2016xyh}, the SM prediction comes with a sizable uncertainty. Recent work using sum rule calculations of hadronic matrix elements quotes $\Delta M_s^\text{SM} = (18.5^{+1.2}_{-1.5}) /$ps~\cite{King:2019lal}. If we instead use the lattice average of hadronic matrix elements from~\cite{Aoki:2019cca} (see also~\cite{Bazavov:2016nty,Dowdall:2019bea}) as well as $|V_{cb}| = (39.9 \pm 1.4) \times 10^{-3}$\footnote{This value is a conservative combination of the inclusive determination of $|V_{cb}|$ quoted in~\cite{PhysRevD.98.030001} and the two recent exclusive determinations from~\cite{Abdesselam:2018nnh} and~\cite{Dey:2019bgc}. The sizable discrepancy between the inclusive and exclusive values is taken into account by rescaling the uncertainty by a factor $2.6$, following the PDG prescription.} we find $\Delta M_s^\text{SM} = (17.7 \pm 1.4) /$ps. This leads to the following bounds at 95\% C.L.
\begin{equation}
 0.85 < \left| \frac{M_{12}}{M_{12}^\text{SM}} \right| < 1.15 ~~~\text{(based on \cite{King:2019lal})}~,\quad 0.87 < \left|\frac{M_{12}}{M_{12}^\text{SM}}\right| < 1.18 ~~~\text{(our evaluation)}~.
\end{equation}
Within our framework, we see that the reasonable agreement of the SM prediction for $B_s$ mixing with the data, which precludes too large a BSM contribution to $M_{12}$, provides an {\em upper} bound on the parameter $v_\Phi$, which cannot therefore be pushed arbitrarily high.\footnote{Note also that the right-hand-side of (\ref{BS mixing}) does not depend on any of the parameters from our $U(1)_X$ gauge sector ({\em i.e.} $g_X, a_e, a_\mu, a_\tau,$ or $a_Y$), but only on the parameters from the mixing and scalar sectors.}
Given the constraints from direct searches place a {\em lower} bound of 3.1 TeV on $v_\Phi$ (see \S \ref{direct search}), we have bounds squeezing the parameter $v_\Phi$ from both sides. For the purposes of this paper, we seek only to show that there is a viable window of parameter space for $v_\Phi$ between these bounds, for which it is sufficient to consider our benchmark point in parameter space (see Table~\ref{tab:charges}). 

\begin{figure}
\centering
\includegraphics[width=1. \textwidth]{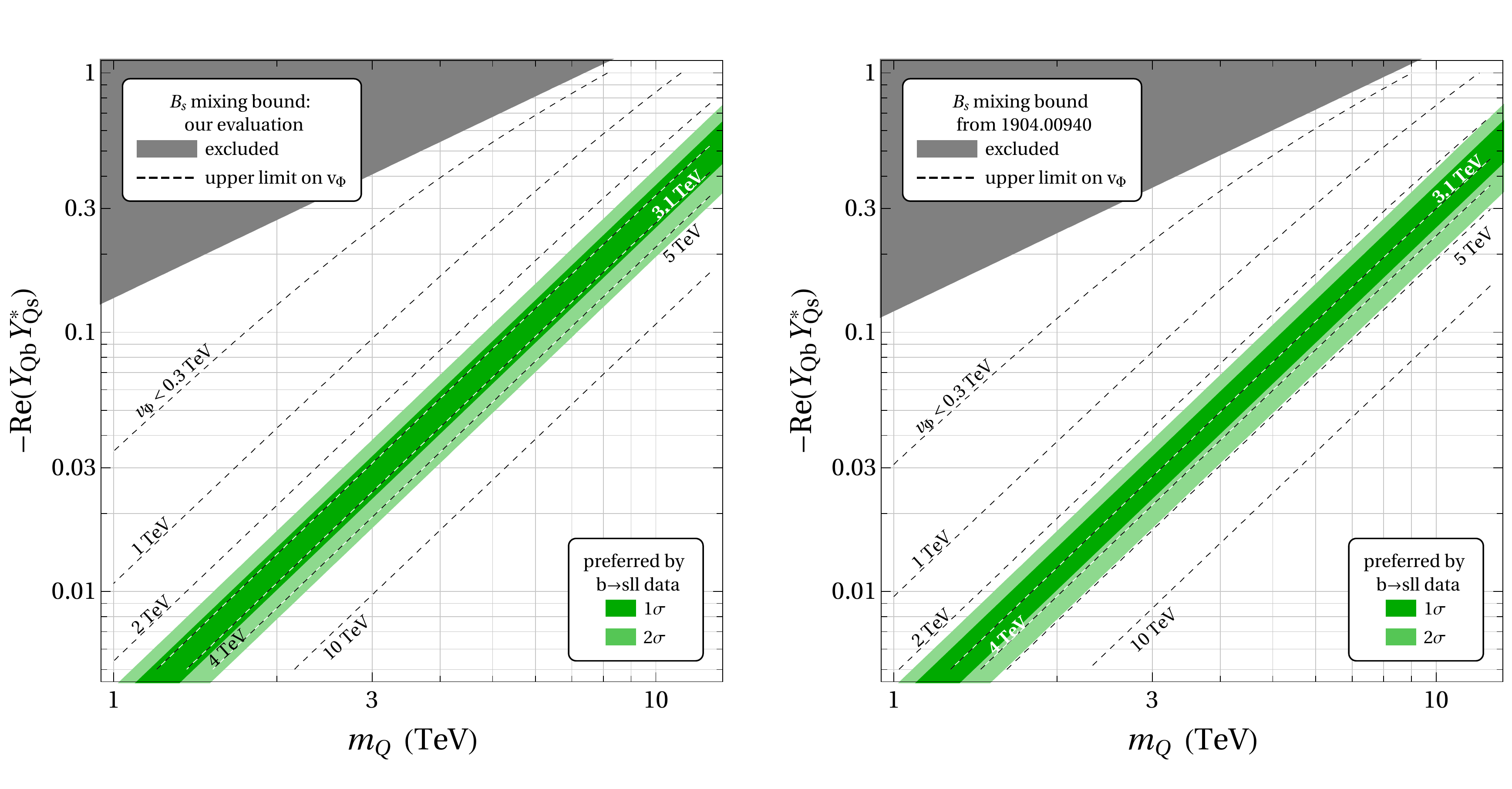}
\caption{Constraints from $B_s$ mixing on $v_\Phi$, in the plane of $\text{Re}(Y_{Qb} Y^*_{Qs})$ {\em vs.} $m_Q$. The shaded green regions show the 1$\sigma$ and 2$\sigma$ best fit regions to the NCBA data, using either the SM prediction of $\Delta M_s$ from~\cite{King:2019lal} (right plot) or using our evaluation (left plot), as discussed in \S \ref{anomalies}. In both plots, the dashed contours show the upper bounds on $v_\Phi$ in TeV coming from $B_s$ mixing. The dark grey regions cannot be made compatible with $B_s$ mixing for any $v_\Phi$. 
\label{fig:mixing}}
\end{figure}

The $B_s$ mixing constraints on $v_\Phi$ at the benchmark point are shown in Fig.~\ref{fig:mixing}, in the plane of $\text{Re}(Y_{Qb} Y^*_{Qs})$ {\em vs.} $m_Q$. In these plots, the shaded green regions show the 1$\sigma$ and 2$\sigma$ best fit regions to the NCBA data, using all the observables, as discussed in \S \ref{anomalies}. Note that the overall normalisation of the Wilson coefficients, which is extracted from the fit to the NCBAs, is proportional to $Y_{Qb} Y^*_{Qs}/m_Q^2$ from (\ref{C9}, \ref{C10}). Hence the green regions correspond to bands with approximately fixed gradient (given the log scale of the plot). Each of the dashed contours then show the upper bounds on $v_\Phi$ coming from $B_s$ mixing. Thus, if we take the central value of the fit to the NCBAS using all observables, we see that the maximum value of $v_\Phi$ is between 3 and 3.5 TeV, depending of the precise $B_s$ mixing bound that is imposed. If we fit the NCBAs at the 1$\sigma$ (2$\sigma$) contours, we can loosen these constraints to $v_\Phi \lesssim 4$\,TeV ($v_\Phi \lesssim 5$\,TeV) respectively.

In the dark grey region in the upper left corner, $B_s$ mixing is never in
agreement with measurements for any value of $v_\Phi$, because the ($v_\Phi$-independent) 1-loop contribution to (\ref{BS mixing}) saturates the bound on its own.

Thus, at the benchmark point the model is tightly squeezed by the combination of constraints on $v_\Phi$ from direct searches and $B_s$ mixing, but there remains a viable window of unexcluded parameter space. Of course, one expects that by deviating from the benchmark point this viable window can be widened (or narrowed); however, a comprehensive analysis of the parameter space is beyond the scope of this paper.

\subsection{Electroweak precision observables}

Here we discuss two important constraints coming from electroweak precision observables (EWPOs), namely measurements of LFU in the $Z$ couplings, and a constraint from the $\rho$-parameter. In general, these constraints from EWPOs arise due to the coupling of the Higgs to $U(1)_X$, and so can be eased by dialling down the value of the parameter $a_Y$ which sets the $\hat{Q}_H$; while $a_Y$ is not essential for fitting the NCBA data, it is, interestingly, the parameter that determines the size of the axial component ({\em i.e.} the Wilson coefficient $C^\alpha_{10}$) of the flavour anomalies, which recall is necessarily flavour universal within our framework. 

\subsubsection{LFU of \texorpdfstring{$Z$}{Z} couplings}

As we discussed in \S \ref{ZZ' mixing}, the $Z$ contains a small admixture of the $U(1)_X$ gauge boson $X$ provided $a_Y\neq 0$, and thus inherits some flavour non-universality in its couplings.  Flavour non-universality in the leptonic decays of the $Z$ are constrained by
the LEP measurement~\cite{PhysRevD.98.030001}
\begin{equation}
R_{\text{LEP}} =0.999\pm 0.003,
\qquad R\equiv\frac{\Gamma(Z\rightarrow e^+e^-)}{\Gamma(Z\rightarrow \mu^+\mu^-)}. \label{LEP}
\end{equation}
In the models we are considering, the ratio of partial widths is
\begin{equation}
R_{\text{model}} = \frac{|g_Z^{e_L e_L}|^2+ |g_Z^{e_R e_R}|^2}{|g_Z^{\mu_L \mu_L}|^2+|g_Z^{\mu_R \mu_R}|^2},
\end{equation}
where $g_Z^{ff}$ is the coupling of the physical $Z$ boson to the fermion pair
$f\bar f$. One can obtain the couplings $g_Z^{ff}$ by first writing down the terms
in the lagrangian which couple the charged leptons to the neutral bosons $B$,
$W^3$, and $X$:
\begin{equation}
\mathcal{L} \supset \sum_{i=1}^3 \left[ \overline{\ell}^i \left(-\frac{1}{2}g\slashed{W}^3-\frac{1}{2}g'\slashed{B} +\frac{1}{2}(2a_i-a_Y)g_X\slashed{X} \right) P_L \ell^i  + \overline{\ell}^i \left(-g'\slashed{B}+(a_i-a_Y)g_X\slashed{X} \right) P_R \ell^i \right]
\end{equation}
We then rotate to the mass basis, and from $X=\sin\alpha_z Z + \cos\alpha_z Z'$, this results in $Z$ couplings that are suppressed by $\sin\alpha_z$. To leading order in $\sin\alpha_z$, we find the couplings are: 
\begin{eqnarray}
\begin{aligned}
g_Z^{e_L e_L}     &= \kappa_L +  \frac{1}{2}(2a_e-a_Y)g_X \sin\alpha_z,  \\
g_Z^{\mu_L \mu_L} &= \kappa_L +  \frac{1}{2}(2a_\mu-a_Y)g_X \sin\alpha_z,  \\
g_Z^{e_R e_R}     &= \kappa_R + (a_e-a_Y)g_X \sin\alpha_z, \\
g_Z^{\mu_R \mu_R} &= \kappa_R + (a_\mu-a_Y)g_X \sin\alpha_z. \\
\end{aligned}
\end{eqnarray}
where
\begin{equation}
\kappa_L\equiv -\frac{1}{2} g\cos\theta_w + \frac{1}{2} g'\sin\theta_w \quad \text{and} \quad \kappa_R\equiv g'\sin\theta_w
\end{equation}
correspond to the couplings in the SM.
We expand $R_{\text{model}}$ to leading order in
$\sin\alpha_z$ to find
\begin{equation}
R_{\text{model}} = 1 + \frac{2g_X\sin\alpha_z (\kappa_L+\kappa_R)(a_e-a_\mu)}{\kappa_L^2+\kappa_R^2}.
\end{equation}
After substituting in (\ref{mass}, \ref{mixing}) for $\sin\alpha_z$ and $m_{Z'}$, and the central experimental values $g=0.64$ and $g'=0.34$, this becomes:
\begin{equation}
R_{\text{model}}= 1 + 1.7957 \left[\frac{(a_\mu-a_e)a_Y}{ v_\Phi^2} \right] m_Z^2.
\end{equation}
Comparison with the LEP limits at the $95\%$ C.L. yields the bounds:
\begin{equation} \label{general Z coupling bound}
\frac{ v_\Phi}{\sqrt{|(a_\mu - a_e)a_Y|}} > \left\{
\begin{aligned}
& 1.73\ \text{TeV}, \quad \text{if}\ (a_\mu - a_e)a_Y > 0, \\
& 1.46\ \text{TeV}, \quad \text{if}\ (a_\mu - a_e)a_Y < 0.
\end{aligned} \right.
\end{equation}
Of course, the bound vanishes when either $a_e=a_\mu$ (in which case the $Z'$, and thus the $Z$ also, couples universally to electrons and muons), or when $a_Y=0$ (in which case the Higgs becomes uncharged, and so there is no $Z-Z'$ mixing).

At our benchmark point (see Table~\ref{tab:charges}) we have $(a_e,a_\mu,a_\tau,a_Y)=(0.59,1,0,0.87)$, and thus $(a_\mu-a_e)a_Y=0.357>0$, so comparing with the first bound in (\ref{general Z coupling bound}) yields the constraint
\begin{equation}
v_\Phi > 1.03\ \mathrm{TeV}.
\end{equation}
This bound is therefore much weaker than that from direct searches computed in \S \ref{direct search}.

\subsubsection{The \texorpdfstring{$\rho$}{rho}-parameter}

The $Z-Z^\prime$ mixing, which occurs at points in our parameter space where $a_Y \neq 0$, also alters the mass of the $Z$ boson away from the SM prediction. In particular, the $\rho$-parameter is no longer equal to unity at tree-level when $a_Y \neq 0$. Global fits to electroweak precision data give the experimental constraint~\cite{PhysRevD.98.030001}
\begin{equation}
\rho = 1.00039 \pm 0.00019.
\end{equation}
The diagonalisation of the neutral gauge boson mass matrix carried out in \S \ref{ZZ' mixing} gives a smaller eigenvalue $m_Z$,\footnote{Of course, the third eigenvalue of (\ref{neutral masses}) is identically zero, corresponding to the massless photon.} which is identified with the $Z$ mass, that satisfies~\cite{Bandyopadhyay:2018cwu}
\begin{equation}
m_Z^2 \cos^2\alpha_z + m_{Z^\prime}^2 \sin^2 \alpha_z = \frac{m_W^2}{\cos^2\theta_w} = \rho m_Z^2,
\end{equation}
where recall the mixing angle is $\sin\alpha_z = \frac{a_Yg_X}{\sqrt{g^2+g'^2}}\left(m_Z/m_{Z^\prime}\right)^2 + \mathcal{O}((m_Z/m_{Z^\prime})^4)$, and the $W$ boson mass is $m_W=vg/2$ as in the SM. This results in the constraint
\begin{equation}
v_\Phi \gtrsim \frac{36\ a_Y m_Z}{\sqrt{g^2+g'^2}}  \simeq a_Y \times 4.4~\mathrm{TeV}
\end{equation}
at the 95\% C.L. limit. At the benchmark point in parameter space ($a_Y = 0.87$) this constraint is in fact very aggressive, implying 
\begin{equation}
v_\Phi > 3.9 ~\mathrm{TeV},
\end{equation}
even stronger than the constraint from direct searches - though still consistent with the 1$\sigma$ best fit to the NCBAs (see Fig.~\ref{fig:mixing}). Of course, this constraint becomes weaker as one deviates away from the benchmark point we have studied in the direction of decreasing $a_Y$. 

\subsection{Neutrino trident production} \label{trident}

The neutrino induced production of a di-lepton pair in the Coulomb field of a heavy nucleus, {\em a.k.a.} neutrino trident production, is known to be an important constraint on models with gauged muon number~\cite{Altmannshofer:2014pba}. 
Muonic tridents, $\nu_\mu \to \nu_\mu \mu\mu$ have been measured at the CCFR experiment~\cite{Mishra:1991bv}. The quoted experimental value for the cross section corresponds to a 95\% C.L. limit of 
\begin{equation}
0.26 < \frac{\sigma(\nu_\mu \to \nu_\mu \mu\mu)_\text{CCFR}}{\sigma(\nu_\mu \to \nu_\mu \mu\mu)_\text{CCFR}^\text{SM}} < 1.38 ~,
\end{equation}
The modifications of the trident cross section at the CCFR experiment coming from the exchange of a virtual $Z^\prime$ gauge boson can be written as~\cite{Altmannshofer:2019zhy}
\begin{equation}
\frac{\sigma(\nu_\mu \to \nu_\mu \mu\mu)_\text{CCFR}}{\sigma(\nu_\mu \to \nu_\mu \mu\mu)_\text{CCFR}^\text{SM}} \simeq \frac{(1+4 \sin^2\theta_W + \Delta g_V)^2 + 1.13\, ( 1 -\Delta g_A)^2}{(1 + 4 \sin^2\theta_W)^2 + 1.13}~,
\end{equation}
where the new physics corrections to the effective couplings $\Delta g_V$ and $\Delta g_A$ are in our model given by
\begin{eqnarray}
 \Delta g_V &=& \left(a_\mu - \frac{3}{4} a_Y\right)\left(2 a_\mu - a_Y\right) \frac{v^2}{v_\Phi^2} ~, \\
 \Delta g_A &=& - \frac{1}{4} a_Y \left(2 a_\mu - a_Y\right) \frac{v^2}{v_\Phi^2} ~. 
\end{eqnarray}
At our benchmark point, the CCFR measurement given above implies the bound 
\begin{equation}
v_\Phi > 0.27  ~\mathrm{TeV}.
\end{equation}
This bound is considerably weaker than the ones from electroweak precision observables and from direct searches at the LHC.

\section{Conclusions} \label{conclusions}

We have proposed a new framework of $Z^\prime$ models based on gauging an almost arbitrary linear combination of the accidental $U(1)$ symmetries of the SM, {\em i.e.} baryon number and individual lepton numbers, as well as global hypercharge. Within this framework of models, the new physics associated with the $Z^\prime$ at the TeV scale respects these global symmetries of the SM, whose breaking is therefore postponed until some even higher energy scale $\Lambda_{\text{LFV}}$ (at which neutrino masses are generated). Such a scenario is hinted at by the recent observations of possible new physics in rare $B$ meson decays, since all these anomalous measurements respect the accidental symmetries of the SM.

The conservation of lepton flavour, in particular, is naturally linked in these models to a violation of lepton flavour {\em universality} at the scale $\Lambda_{\text{LFUV}}$ at which the $Z^\prime$ resides (since non-universal lepton charges are required to align the weak and mass eigenbases for charged leptons, which protects lepton flavour). Such a $Z^\prime$ model therefore offers a tempting explanation of the neutral current $B$ anomaly data, in which lepton flavour universality is observed to be violated between $e$ and $\mu$ in $b\rightarrow s$ transitions, and we explore this possibility. The requisite quark flavour violation is introduced through a heavy vector-like quark state.

In the rest of the paper we explored the phenomenology of these models. Our analysis of the parameter space is not comprehensive. Rather, we were content to point out some interesting features that are common to all these models, such as the freedom to add a flavour-universal axial component to the $B$ anomaly fit. We then made a number of well-motivated assumptions to restrict the parameter space to a region of particular relevance to explaining the neutral current $B$ anomalies, and by performing a global fit in this region we extracted a benchmark point in our space of models, at which to examine the phenomenology more carefully. At this benchmark point the $Z^\prime$ couples to both left-handed and right-handed electrons and muons. We have shown that the model is consistent at this benchmark point with bounds from direct LHC searches, $B_s$ mixing, electroweak precision observables, and neutrino trident production. 

Finally, in an Appendix we discuss how neutrino masses can be generated in such a setup, which involves a rather detailed construction of a dark sector. One can thence relate neutrino masses to the high energy scale $\Lambda_{\text{LFV}}$ at which the accidental symmetries of the SM are eventually broken within our framework. This gives an estimate $\Lambda_{\text{LFV}}\sim 10^5$ TeV, which is indeed much heavier than the scale $\Lambda_{\text{LFUV}}\sim 1$ TeV of new physics associated with the $Z^\prime$, and so is therefore consistent with our original hypothesis that $\Lambda_{\text{LFV}} \gg \Lambda_{\text{LFUV}}$.

An important next step would be to carry out a more comprehensive study of the phenomenological constraints on the parameter space of these models; for example, one might like to incorporate constraints from electroweak precision observables such as the $\rho$-parameter into the global fit to the flavour anomaly data, since we have seen that such precision observables do provide important constraints on our parameter space, in particular on the value of the parameter $a_Y$. 

Another promising future direction is to explore an alternative `benchmark scenario' to the one we considered in this paper, in which the couplings of the $Z^\prime$ to light quarks are set to zero, thereby significantly loosening up the constraints from direct searches at the LHC. This requires $a_Y=0$ and $a_e+a_\mu+a_\tau=0$, and so leaves two independent parameters to scan over in our fit to the flavour anomaly data. Given $a_\tau$ is essentially unconstrained phenomenologically, one would most likely want to freely scan over values of $a_e$ and $a_\mu$ (setting $a_\tau=-a_e-a_\mu$). Setting $a_Y=0$ also has the consequence of removing the tree-level $Z-Z^\prime$ mixing, and so would also relax the bounds from electroweak precision observables (such as that coming from the $\rho$-parameter), which we found to be stringent constraints for large values of $a_Y$. We therefore expect the phenomenology to be much more open in this region than in the benchmark we chose to study in this paper. Of course, taking this limit has significant implications for the flavour anomalies, since it also removes the possibility of any axial component in the flavour anomaly explanation. This scenario gives a simple illustration of the interesting interplay between different experimental constraints and the character of the flavour anomalies within our framework of gauging the accidental symmetries of the SM.

\section*{Acknowledgements}

We would like to thank Ben Allanach, David Marzocca, Scott Melville, and Tevong You for helpful discussions. We are especially grateful to David Marzocca for his computation of the bounds on our model (at the benchmark point in parameter space) coming from $pp\rightarrow \mu^+\mu^-$ and $pp\rightarrow e^+ e^-$ direct searches, described in \S \ref{direct search}. We also thank other members of the Cambridge Pheno Working Group for helpful comments. JD is supported by The Cambridge Trust and by the STFC consolidated grant ST/P000681/1.
The research of WA is supported by the National Science Foundation under Grant No. PHY-1912719.
WA acknowledges support by the Munich Institute for Astro- and Particle Physics (MIAPP) of the DFG cluster of excellence ``Origin and Structure of the Universe''.

\appendix

\section{Neutrino masses and the dark sector} \label{neutrinos}

Recall that a primary goal of this paper was to extend the SM in a way that preserves its global symmetries, in particular $U(1)_B$ and each individual lepton number $U(1)_{L_i}$, which are experimentally tested to very high precision. From this premise, we arrived at the four-parameter family of anomaly-free $U(1)_X$ charge assignments recorded in (\ref{charges}), for which the alignment of the charged lepton mass basis with the weak eigenbasis leads to a natural protection of lepton numbers, while at the same time predicting lepton flavour universality violation.

Nonetheless, we know that the lepton number symmetries cannot, ultimately, be exact symmetries of Nature, due to one important observation: the measurement of neutrino oscillations. Within the framework we have put forward in this paper, there is (by design) no lepton flavour violation in charged leptons, at least at the renormalisable level to which we have worked so far, and so we are led to suppose that all the observed lepton flavour violation (as parametrized, for example, by the PMNS matrix elements) must originate from the neutrino sector. Recall that in addition to the left-handed weakly-interacting neutrinos of the SM there are three right-handed SM singlet states in our setup, which we introduced to `soak up' the $U(1)_X^3$ and gravitational anomalies; the natural interpretation of these dark states is to identify them as right-handed neutrinos, which are charged only under the $U(1)_X$ gauge symmetry.

The measurement of neutrino mass-squared differences of the order $10^{-3}$ eV$^2$ does not, however, point us ambiguously towards an energy scale $\Lambda_{\text{LFV}}$ (to use the terminology introduced in \S \ref{intro}) at which the lepton number symmetries become broken, because such a cut-off scale is highly dependent on the physical mechanism which gives rise to the neutrino mass terms.
However, with the charge assignments presented in (\ref{charges}), we in fact find a neutrino mass sector that is in tension with our hypothesis that $\Lambda_{\text{LFV}}$ resides much higher than the TeV scale ($\Lambda_{\text{LFUV}}$) associated with the $Z^\prime$ boson and corresponding LFUV effects, where $\Lambda_{\text{LFUV}} \sim v_\Phi$. In this Section we shall explain why there is such a tension in the neutrino sector, before sketching how the dark sector of the model can be altered to resolve this tension, and thus give a compelling account of neutrino masses and the eventual breaking of lepton number symmetries.

The reason for the apparent tension is as follows. Firstly, one can use the three dark states $\nu_R^i$ to write down a diagonal matrix of renormalisable Yukawa couplings for neutrinos, of the form\footnote{Of course, even if the Yukawa couplings $y_i$ were tiny, these terms on their own cannot explain the neutrino mass sector, because the PMNS matrix features large mixing angles.} 
\begin{equation}
\mathcal{L}_{\text{Yuk}}=\sum_{i=1}^3 y_i \overline{\ell}^i_L H \nu_R^i.\label{yukawa}
\end{equation}
We can also write down Majorana mass terms involving the SM singlet states $\nu_R^i$, possibly with insertions of the scalar field $\Phi$ to soak up the $U(1)_X$ charge, depending on the particular values of $a_e$, $a_\mu$, and $a_\tau$. For example, in the benchmark case that we defined in \S \ref{fits} (and have studied phenomenologically in \S \ref{pheno}), we can write down the following operators
\begin{equation}
\mathcal{L}_{\text{Maj}} \supset \alpha(\overline{\nu}^{2}_R)^c\ \Phi^{\ast}\nu^3_R + \beta(\overline{\nu}^{3}_R)^c\ \Phi^{\ast}\nu^2_R + \gamma M (\overline{\nu}^{3}_R)^c\ \nu^3_R, \label{majorana}
\end{equation}
where $M$ indicates a mass scale which is {\em a priori} unrelated to any other scale in the theory, and the couplings $\alpha$, $\beta$, and $\gamma$ are all dimensionless coefficients. After spontaneous breaking of $U(1)_X$, the dimension four operators in (\ref{majorana}) lead to Majorana mass terms with coefficients set by the scale $v_\Phi$, which we know is of order a few TeV.

These effective Majorana operators break lepton flavour symmetries, in this case $U(1)_\mu$ and $U(1)_\tau$, at the scale of $v_\Phi$. Now, if this lepton number violation were confined to the dark sector states this would not pose the problem, since from the point of view of the SM fields the SM's accidental global symmetries would remain intact up to the higher scale $\Lambda_{\text{LFV}}$. But that is not the case, because the dark states $\nu_R^i$ necessarily interact with the charged leptons $\ell_i$ through the diagonal Yukawa interactions at a low energy scale,\footnote{One rather unsatisfying resolution to this problem is that there is in fact a discrete $\mathbb{Z}_2$ symmetry which bans these Yukawa interactions.} and we therefore have no natural mechanism for preserving lepton flavour symmetries in the SM sector up to the high scale $\Lambda_{\text{LFV}}$. Thus, the tension arises precisely because of the interplay between the lepton flavour violating Majorana interactions with the Yukawa couplings which couple the dark sector to the SM.

\subsection{Alternative dark sectors}

It is possible to resolve this tension by exploring alternative dark sectors.\footnote{We note in passing that the idea of using extra chiral states which `soak up' gauge anomalies to serve as dark matter has been explored before in the context of gauging $U(1)_{B-L}$~\cite{Nakayama:2011dj,Lindner:2011it,Ibe:2011hq,Lindner:2013awa}. In this paper, we do not consider dark matter phenomenology. } To see how this can be done, recall that the dark sector states $\nu_R^i$ were introduced to `soak up' the non-vanishing gauge (and gauge-gravity) anomalies. We introduced {\em three} dark states because that was sufficient to guarantee cancellation of all anomalies for {\em any} choice of the four rational coefficients $(a_e,a_\mu,a_\tau,a_Y)$ - see equations (\ref{dark anomalies}, \ref{dark charges}). 

We here remark in passing that whatever extra chiral states we introduce the `soak up' any field theory anomalies are better off being dark, for two reasons. Even if we added SM non-singlet states that were chiral only under $U(1)_X$, these states would be on the one hand dangerous from the point of view of phenomenology, since they would acquire masses set by $v_\Phi \sim 3$ TeV multiplied by Yukawa-like dimensionless couplings (and so might easily have masses of 1 TeV or lower), for which the experimental bounds on, say, coloured states are strong (see for example recent bounds from gluino searches \cite{Aad:2019tcc}), and on the other hand less useful from the point of view of soaking up anomalies, since (for example) an $SU(3)$ triplet would always appear in (\ref{dark anomalies}) with a multiplcity factor of 3. If the added states were completely chiral under the SM the situation is worse still, because their masses would then be set by the Higgs vev $v$, which is very problematic for LHC phenomenology even if such states were colourless. 

However, as we noted above (see footnote 12), the assignment of dark charges in (\ref{dark charges}) does not provide the {\em only} solution to the Diophantine equations (\ref{dark anomalies}). As we there observed, for {\em particular values} of $(a_e,a_\mu,a_\tau)$ there may be other solutions for the dark charges which soak up the anomalies; generically, these other solutions will not allow Yukawa couplings of the form (\ref{yukawa}), and in this way may circumvent the tension presented above. We will at times refer to these other solutions as `non-trivial solutions' to (\ref{dark anomalies}), since the choice made in (\ref{dark charges}) is trivial from the number theory perspective. Indeed there are even non-trivial solutions with only two dark states,\footnote{For example, with only two dark states $\nu_R^1$ and $\nu_R^2$, there are anomaly-free solutions with $(a_e,a_\mu,a_\tau,-\hat{Q}_{\nu^1},-\hat{Q}_{\nu^2})$ equal to (any permutation of) the sets $(7,8,-9,-1,-5)$, $(9,-2,-10,-4,7)$, {\em etc}.} and certainly there will be many more if we allow say four or more dark states. The general solutions presented in Ref.~\cite{Costa:2019zzy} can be adapted to generate solutions to this pair of equations at will, for any number (greater than one) of chiral dark states.

To furnish us with a concrete example of such an alternative dark sector, consider the benchmark case we set up in \S \ref{fits}. By normalising to $a_\mu=1$ and setting $a_\tau=0$ for simplicity,\footnote{We emphasize that the choice $a_\tau=0$ was really just to simplify the discussion and the notation, rather than simplify the physics; the value of $a_\tau$ does not affect any of the relevant phenomenological bounds computed in \S \ref{pheno}.} we found the best-fit point \begin{equation}
(a_e,a_\mu,a_\tau,a_Y)=(0.59,1,0,0.87),
\end{equation}
by performing a global fit to the NCBA data (see Fig.~\ref{fig:fits}), where we assumed the anomalies in (\ref{dark anomalies}) were soaked up by states with charges $\hat{Q}_{\nu^1}=0.59$, $\hat{Q}_{\nu^2}=1$, and $\hat{Q}_{\nu^3}=0$ ({\em i.e.} exploiting the `trivial solution' to the anomaly equations).
But now consider an anomaly-free model with an alternative dark sector, also with three dark chiral fermion states, in which
\begin{equation}
(a_e,a_\mu,a_\tau,\hat{Q}_{\nu^1},\hat{Q}_{\nu^2},\hat{Q}_{\nu^3})=\left(\frac{3}{5},1,\frac{1}{20},\frac{1}{4},\frac{7}{20},\frac{21}{20} \right),
\end{equation}
corresponding to a non-trivial rational solution to the anomaly cancellation equations (\ref{dark anomalies}).
This alternative anomaly-free charge assignment coincides (very nearly) with the benchmark case when restricted to the SM fields, and so shares the same phenomenology. The dark sector is however very different, which leads to a different story concerning neutrino masses, which we shall soon explain.

The crucial point is that with these dark sector charges we can no longer write down renormalisable Yukawa couplings of the form (\ref{yukawa}) which couple the dark sector to the SM fermions. At the renormalisable level, the neutrinos are now strictly massless, and we must pass to the SMEFT to explain the origin of neutrino masses.

Before we do so, we should first give masses to the dark states. While massless dark fermions of this kind are not in conflict with data from collider experiments, they would have profound consequences for the cosmological evolution of the Universe. To avoid a detailed analysis of the cosmological constraints, one can make the dark fermions heavy by introducing a pair of extra dark scalars transforming in representations
\begin{equation} \label{dark scalars 1}
\chi_1 \sim \left(1, 1, 0, \frac{3}{5}\right), \quad \chi_2 \sim \left(1, 1, 0, \frac{21}{10}\right)
\end{equation}
of $SU(3)\times SU(2)_L \times U(1)_Y \times U(1)_X$,
which are both assumed to acquire non-vanishing vevs upon spontaneous breaking of the $U(1)_X$ gauge symmetry. One may then write down the following dimension four operators involving pairs of dark fermions and a single dark scalar
\begin{equation}
\mathcal{L}_{\text{dark}}\supset \mathcal{M}^\text{dark}_{ij}\; (\overline{\nu}_R^i)^c \chi_a \nu_R^j, \qquad \mathcal{M}^\text{dark}_{ij}\sim
\left( \begin{array}{ccc}
0 & \times & 0 \\
\times & 0 & 0 \\
0 & 0 & \times \\
\end{array}\right),
\end{equation}
where here $\chi_a$ indicates either of the dark scalars in (\ref{dark scalars 1}). $\mathcal{M}^\text{dark}$ is a rank-3 mass matrix and so leads to three non-zero masses for the dark fermions, all at the scale of $U(1)_X$ breaking ({\em i.e.} the TeV scale).

\subsection{Neutrino masses: estimating the scale of lepton flavour violation}

If for simplicity we take $a_Y=9/10$ (which is close to the best-fit value of $a_Y=0.87$ obtained in \S \ref{fits}), the left-handed lepton doublets have the rational charges 
\begin{equation}
\hat{Q}_{\ell^1}=3/20, \qquad \hat{Q}_{\ell^2}=11/20, \qquad \text{and} \quad \hat{Q}_{\ell^3}=-2/5,
\end{equation}
which are, by construction, numerically almost equal to the charges in the benchmark case as recorded in Table~\ref{tab:charges}.

One can show that mass terms involving left-handed neutrinos first appear at dimension six in the SMEFT, due to lepton flavour violating operators of the form
\begin{equation}
\frac{c_{ij}}{\Lambda_{\text{LFV}}^2} \ell_i H \ell_j H \chi_a,
\end{equation}
where again $\chi_a$ here denotes any dark scalar charged under $U(1)_X$. After spontaneous breaking of $U(1)_X$, these dimension six operators reduce to the familiar dimension five Weinberg operators. Note the appearance of the scale $\Lambda_{\text{LFV}}$ in this EFT expansion; recall $\Lambda_{\text{LFV}}$ reflects the scale of new physics which may {\em break} the global symmetries of the SM, in particular lepton flavour, in other words the scale at which our $Z^\prime$ model breaks down. The lower scale $\Lambda_{\text{LFUV}}\sim 1$ TeV is the cut-off scale at which the SMEFT breaks down, being resolved at short-distances by our $Z^\prime$ model. 

We can populate the mass matrix $c_{ij}$ up to a single zero, {\em viz.}
\begin{equation}
c_{ij}\sim \left( \begin{array}{ccc}
\times & \times & \times \\
\times & \times & \times \\
\times & \times & 0 \\
\end{array}\right),
\end{equation}
if we introduce three more dark scalars transforming in representations
\begin{equation} \label{dark scalars 2}
\chi_3 \sim \left(1, 1, 0, \frac{1}{5}\right), \quad \chi_4 \sim \left(1, 1, 0, \frac{3}{4}\right), \quad \chi_5 \sim \left(1, 1, 0, \frac{23}{20}\right),
\end{equation}
which is known to be a sufficiently dense texture to accommodate present data on neutrino masses and mixings. We emphasize that our construction of a viable dark sector in this Appendix is intended as a proof of principle, and we do not intend for this setup to be understood as being in any way a `minimal choice'.\footnote{For example, one could rescale the $U(1)_X$ charge of the scalar field $\Phi$ introduced in the main text to play the role of one of the five scalars $\chi_a$ discussed in this Appendix, thereby eliminating one scalar, as long as one shifts the gauge coupling $g_X$ and the charge $\hat{Q}_Q$ of the vector-like quark accordingly.} 

Finally, we can now estimate the higher cut-off scale $\Lambda_{\text{LFV}}$ associated with lepton flavour violation, using the scale of neutrino masses. At the level of na\"ive dimensional analysis we require 
\begin{equation}
\frac{v^2 v_\Phi}{\Lambda_{\text{LFV}}^2} \sim m_\nu \approx 10^{-13} ~\text{TeV} \quad \implies \quad \Lambda_{\text{LFV}} \gtrsim 10^5 ~\text{TeV},
\end{equation}
which sure enough far exceeds the TeV scale associated with the $Z^\prime$ and LFUV effects such as the measurements of $R_{K^{(\ast)}}$.

\section{Details of the fit to rare $B$ decay data} \label{app:fit}

In this Appendix we provide details about the fit to rare $B$ meson decay data that we perform to identify the preferred parameter space of our model. 

We carry out two fits in the 3-dimensional parameter space of Wilson coefficients $C_9^\mu$, $C_9^e$ and $C_{10}^\mu = C_{10}^e \equiv C_{10}$ using \texttt{flavio}~\cite{Straub:2018kue}. In fit 1, we include the measurements of LFU ratios $R_K$ and $R_{K^*}$ from LHCb~\cite{Aaij:2017vbb,Aaij:2019wad} and Belle~\cite{Abdesselam:2019wac,Abdesselam:2019lab}, the combination of the $B_s \to \mu \mu$ branching ratio from~\cite{Aebischer:2019mlg}, that includes data from LHCb, CMS, and ATLAS~\cite{Chatrchyan:2013bka,CMS:2014xfa,Aaij:2017vad,Aaboud:2018mst}, as well as measurements of the branching ratios of the inclusive decays $B \to X_s \ell\ell$ from BaBar~\cite{Lees:2013nxa} and Belle~\cite{Iwasaki:2005sy}. We find a compact region of parameter space that is compatible with the considered data and we approximate the best fit region by a multivariate Gaussian. The corresponding central values for the Wilson coefficients, their uncertainties and the correlation matrix are 
\begin{equation}
 \begin{matrix}
  C_9^\mu = 0.37 \pm 0.76 \\
  C_9^e = 1.13 \pm 0.78 \\
  C_{10} = 0.63 \pm 0.29 \\
 \end{matrix} ~, \qquad
 \rho = \begin{pmatrix}
         1 & 0.95 & 0.34 \\ 0.95 & 1 & 0.29 \\ 0.34 & 0.29 & 1
        \end{pmatrix}~, \qquad \text{``clean observables''}~.
\end{equation}
In fit 2, we include in addition measurements of the branching ratios of the exclusive semileptonic decays $B^\pm \to K^\pm \mu\mu$, $B^0 \to K^0 \mu\mu$, $B^0 \to K^{*0}\mu\mu$, $B^\pm \to K^{*\pm}\mu\mu$, $B_s \to \phi \mu\mu$ and $\Lambda_b \to \Lambda \mu\mu$, as well as all available measurements of CP averaged angular coefficients in these decays from LHCb, CMS, and ATLAS~\cite{Aaij:2014tfa,Aaij:2014pli,Aaij:2015oid,Aaij:2015esa,Aaij:2015xza,Khachatryan:2015isa,Aaij:2016flj,CMS:2017ivg,Aaij:2018gwm,Aaboud:2018krd}. The best fit region is given by
\begin{equation}
 \begin{matrix}
  C_9^\mu = -0.71 \pm 0.24 \\
  C_9^e = 0.12 \pm 0.35 \\
  C_{10} = 0.44 \pm 0.19 \\
 \end{matrix} ~, \qquad
 \rho = \begin{pmatrix}
         1 & 0.65 & 0.19 \\ 0.65 & 1 & 0.19 \\ 0.19 & 0.19 & 1
        \end{pmatrix}~, \qquad \text{``all observables''}~.
\end{equation}
Using eqs.~(\ref{eq:C1}) and~(\ref{eq:C2}), the best fit regions in the parameter space of the Wilson coefficients can be mapped onto the model parameters $a_e$, $a_Y$, and $C$. The result is shown in Fig.~\ref{fig:fits} for the $a_e$ {\em vs.} $a_Y$ plane, profiling over the normalisation factor $C$.

\bibliographystyle{JHEP-2}
\bibliography{articles}
\end{document}